\documentclass[]{aa}
\usepackage{graphicx}
\usepackage{natbib}
\usepackage{txfonts}
\usepackage{aalongtable}
\bibpunct{(}{)}{;}{a}{}{,}

\begin{document}

\title{Long-term monitoring of $\theta^1$ Ori\,C: the spectroscopic orbit and an improved rotational period}

\titlerunning{Long-term monitoring of  $\theta^1$ Ori\,C}

\author{O. Stahl
  \inst{1}
  \and  G. Wade
  \inst{2}
  \and V. Petit
  \inst{3}
  \and B. Stober
  \inst{4}
  \and L. Schanne
  \inst{5}}

 \institute{ZAH, Landessternwarte K\"onigstuhl, 69117 Heidelberg, Germany.
  \email{O.Stahl@lsw.uni-heidelberg.de}
  \and 
  Dept.\ of Physics, Royal Military College of Canada, PO Box 17000, Station Forces, Kingston, ON, Canada K7K 7B4
  \and
D\'epartement de physique, g\'enie physique et optique, Centre de recherche en astrophysique du Qu\'ebec, Universit\'e Laval, Qu\'ebec (QC) G1K 7P4
  \and
   Nelkenweg 14, 66791 Glan-M\"unchweiler, Germany
  \and
  Hohlstra{\ss}e 19, 66333 V\"olklingen-Ludweiler, Germany}

\offprints{O. Stahl} 

\date{Received / Accepted} \abstract 
     {The young O-type star $\theta^1$\,Ori\,C, the brightest star of
       the Trapezium cluster in Orion, is one of only two known
       magnetic rotators among the O stars.  However, not all
       spectroscopic variations of this star can be explained by the
       magnetic rotator model. We present results from a long-term
       monitoring to study these unexplained variations and to improve
       the stellar rotational period.}
     {We want to study long-term trends of the radial velocity of $\theta^1$\,Ori\,C, to search for unusual changes,
       to improve the established rotational period and to check for possible period changes.} 
     {We combine a large set of published spectroscopic data with new observations and analyze the spectra in a homogeneous way.
       We study the radial velocity from selected photo-spheric lines and determine the equivalent width of the
       H$\alpha$ and He{\sc ii}\,$\lambda4686$ lines.} 
     {We find evidence for a secular change of the
     radial velocity of $\theta^1$\,Ori\,C that is consistent with the published
     interferometric orbit. We refine the rotational period of $\theta^1$\,Ori\,C
     and discuss the possibility of detecting period changes 
     in the near future.}
     {} 
     
\maketitle

\section{Introduction}

The young O star $\theta^1$\,Ori\,C, the brightest star of the
Trapezium cluster in Orion, is one of the only two O-type stars with detected
magnetic fields \citep[the other is
  HD\,191612, cf.][]{2006MNRAS.365L...6D}.  Regular variations of the
H$\alpha$ line with a period of 15.4 days were discovered from
equivalent width measurements by \citet{1993A&A...274L..29S}. Later,
this same period was also detected in e.g.\ UV spectral lines
\citep{1994ApJ...425L..29W} and the X-ray flux
\citep{1997ApJ...478L..87G}.

The magnetic field of $\theta^1$\,Ori\,C, which also varies according
to the 15.4 day period, was first detected by
\citet{2002MNRAS.333...55D} and later studied in more detail by
\citet{2006A&A...451..195W}.  For a detailed spectroscopic analysis of
the star see \citet{2006A&A...448..351S}.

The observations are explained conceptually by
the magnetically-confined wind shock (MCWS) model, originally proposed by
\citet{1997ApJ...485L..29B}. In this
model, a dipolar magnetic field confines the outflowing radiatively-driven stellar wind, which is
channeled toward the magnetic equator where it generates a strong shock. The resulting circumstellar
plasma is forced to rotate with the star, generating periodic variability of the emitted optical, UV and X-ray fluxes. 
This model has more recently been extended using MHD simulations
by \citet{2002ApJ...576..413U} and \citet{2005ApJ...628..986G}, who have investigated the
stability and dynamics of this phenomenon.

The period of $\theta^1$\,Ori\,C was determined to be 15.422 $\pm$ 0.002
days by \citet{1996A&A...312..539S} and later revised to 15.426 $\pm$
0.002 by \citet{stahletaleso}. However, most publications
\citep[e.g.][]{2006A&A...451..195W} use the older period value of
15.422 days. The difference in both periods has now accumulated to a
phase difference of about 0.1, which is quite significant. Observations obtained
at the current epoch should therefore be able to distinguish between these two periods.

In addition to the strict periodicity, $\theta^1$\,Ori\,C also shows
additional variations, which are probably not periodic, or have
unknown periods e.g.\ the spectral type variations reported by
\citet{1981ApJ...243L..37W} or the radial velocity variations found by
\citet{1996A&A...312..539S}.  Also, $\theta^1$\,Ori\,C is in fact a
multiple system and interferometric measurements recently propose a
long orbital period of more than ten years
\citep{2007A&A...466..649K,2008ApJ...674L..97P}.

The published radial-velocity measurements have been analyzed 
by \citet{2002AstL...28..324V}, but more data are available.

Long-term monitoring is necessary to study these variations.  We
therefore collected all available published spectra of
$\theta^1$\,Ori\,C and obtained new observations to study long-term
trends, to search for unusual variations and to improve the
determination of the rotational period.

\section{Observations}

For the study of the long-term variations, we primarily used published
echelle observations: The Heros data published by
\citet{stahletaleso}, complemented by a few other observations
obtained with the same instrument; the Feros observations published by
\citet{2000A&A...363..585R}; the MuSiCoS spectra published by
\citet{2006A&A...451..195W}; a few ESPaDOnS spectra \citep{petit2008}
and two spectra extracted from the Elodie archive
\citep[][http://atlas.obs-hp.fr/elodie/]{2004PASP..116..693M}. All of
these observations cover a large spectral range with high resolving
power.

In addition, we have been obtained a few spectra with amateur
instruments. For these observations we used the spectrograph
Lhires\,III, which is available from http://www.shelyak.com/, attached
to Celestron 14$^{\prime\prime}$ Schmidt-Cassegrain telescopes at
private observatories. The detector for most observations was a CCD
with 2184 $\times$ 1472 pixels (used with 2 $\times$ 2 binning) with a
pixel size of 6.8 $\mu$m. A grating with 1200 lines/mm was used for
most observations (Lhires\,III,1), resulting in a spectral resolution
of about 1.0 \AA\@. A few spectra were obtained with a grating with
2400 lines/mm, resulting in a higher resolution of 0.5
\AA\ (Lhires\,III,2). One spectrum was obtained with the 2400 lines/mm
grating, but with another detector with 1536 $\times$ 1024 pixels with
9 $\mu$ pixel size (Lhires\,III,3). All of these spectra were reduced
with ESO-Midas, using standard procedures. The wavelength calibration
was performed using a Neon lamp. The spectra of $\theta^1$ Ori\,C show
strong nebular lines from the Orion nebula. These lines were used to
verify the wavelength solution. The Lhires observations cover a
relatively small spectral range around H$\alpha$ and were used only
for the equivalent width determination of H$\alpha$. A summary of the
data used is given in Table~\ref{tab:inst}.

\begin{table*}
\caption{Summary of instrumentation used for this study. Most of the
  spectra (except the Lhires spectra, have been partly published. The
  signal-to-noise ratio for the echelle data is strongly dependent on
  wavelength, but above 100 in the spectral ranges used for most of
  the spectra.}
\begin{tabular}{llll}
\hline
Instrument & resolution [$\lambda/\Delta\lambda$] & wavelength range [\AA]&  JD - 2\,400\,000 \\
\hline
Flash & 20\,000 & 4\,000 -- 6\,800 & 48\,822 -- 49\,333 \\
Heros & 20\,000 & 3\,450 -- 5700, 5\,800 -- 8\,625 & 49\,759 --  50\,727 \\
Feros & 48\,000 & 3\,700 -- 9\,200 &  51097 -- 51394,  53740 \\
Elodie & 42\,000 & 4\,000 -- 6\,800 & 50030, 53329 \\
Musicos & 35\,000 &  4\,480 -- 6\,620 &51577 -- 51608 \\ 
Espadons & 65\,000 & 3\,690 -- 10\,485 & 53744,  54169, 54456 -- 54457 \\
Lhires\,III,1 & 6\,000 & 6\,350 -- 6\,900 & 54505 -- 54521 \\
Lhires\,III,2 & 12\,000 & 6\,500 -- 6\,700 & 54527 -- 54544\\ 
Lhires\,III,3 & 12\,000 & 6\,500 -- 6\,700 & 54505    \\
\hline
\hline
\end{tabular}
\label{tab:inst}
\end{table*}

\section{Radial velocity variations}

\begin{figure}
\begin{center}
\includegraphics[height=8.8cm,angle=-90]{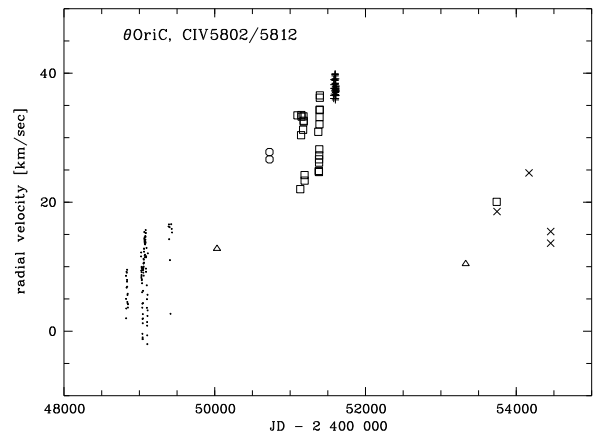}
\end{center}
\caption{Radial velocity of C{\sc iv}\,$\lambda\lambda$5801, 5812
  versus Julian Date. The symbols denote different instruments. $\bullet$: Flash, $\circ$: Heros, $\Box$: Feros, $\bigtriangleup$: Elodie,
  $+$: Musicos, $\times$: Espandons}
\label{fig:vrad}
\end{figure}

The radial velocity variations of $\theta^1$\,Ori\,C have been studied
by various authors \citep[e.g.,][]{2002AstL...28..324V}, but with
ambiguous results. It appears likely that the variations reflect the
15.4 day rotational modulation, as well as mysterious, shorter- and
longer-term variations.  However, the published radial velocities show
significant scatter.  In light of the interferometric orbits published
by \citet{2007A&A...466..649K} and \citet{2008ApJ...674L..97P} with a
period of more than ten years, a re-analysis of the radial velocities
seems warranted.  As most spectral lines vary significantly with the
15.4 day period, we decided to use only the C{\sc iv} line doublet at
$\lambda\lambda$5801.51, 5812.14 \AA, the He{\sc
  ii}\,$\lambda$5411.424 line and the O{\sc iii}\,$\lambda$5592.37
line for radial velocity studies. These lines seem to be more weakly
affected by the rotational modulation \citep{1996A&A...312..539S} and
as a group they provide consistent results.  The lines were modeled by
fitting Gaussians to the line profiles, which matches the lines very
well. The result is reported in Tab.\,\ref{tab:vrad2} and plotted for
the C{\sc iv} lines in Fig.\,\ref{fig:vrad}. A closer analysis shows
small, but significant systematic offsets of the order of a 2--3 km
s$^{-1}$ between different lines. These offsets are probably due to
blends or atmospheric stratification.  All lines show the same
variability pattern, however.

\begin{figure}
\begin{center}
\includegraphics[width=8.8cm,angle=0]{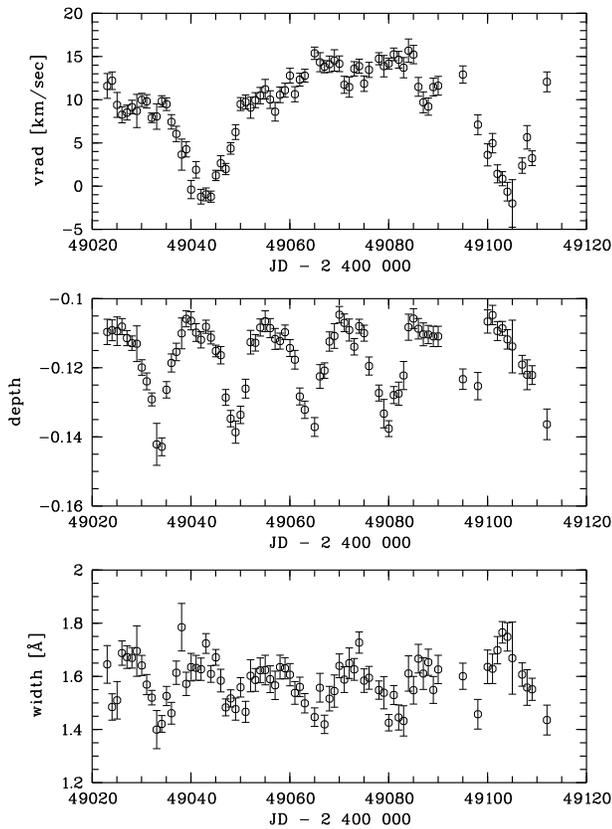}
\end{center}
\caption{Average radial velocity, line depth and FWHM, together with
  the 1\,$\sigma$ error bars, of the C\,{\sc IV}$\lambda$5801, 5812
  lines for part of the time covered.  The radial velocity shows
  strong variations, but not with the 15.4\,d period.  The line depth
  and width are clearly variable with the 15.4\,d period. }
\label{fig:para}
\end{figure}

As can been seen in Fig.\,\ref{fig:vrad}, a large scatter on short
time scales is obvious. These variations have already been detected by
\citet{1993A&A...274L..29S}. The scatter is partly due to variations
with the rotational period, but primarily it is caused by occasional
variations on other timescales (cf.\@ Fig.\,\ref{fig:para}). From
Fig.\,\ref{fig:para}, a period of about 60 days seems possible.
However, a period analysis of the radial velocities does not show a
significant peak near this period. In the period range below 100 days,
only the 15.4 day period is significant in the radial-velocity data.
Therefore, the rapid variations in radial velocity are probably not
periodic, and we speculate that they may be due to (stellar)
atmospheric effects.

\begin{figure}
\begin{center}
\includegraphics[height=8.8cm,angle=-90]{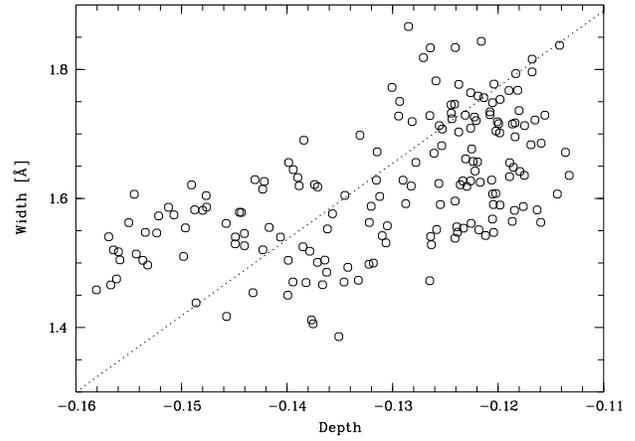}
\end{center}
\caption{Correlation of FWHM and line depth of C\,{\sc
    IV}$\lambda$5801, 5812.}
\label{fig:corr}
\end{figure}

The Gaussian fit to the line also measures the line width (FWHM) and
depth of the lines. Both quantities are strongly variable on short
timescales (by about $\pm$30\%), but without any obvious long-term
trend. In contrast to the radial velocities, the variations in width
and depth are mostly due to rotational modulation
(cf.\@ Fig.\,\ref{fig:para}).  The width of the lines and their depth are
clearly correlated.  The lines are deeper when they are narrower, see
Fig.\,\ref{fig:corr}.  The equivalent widths are therefore less
variable, by only about $\pm$10\%.

In Fig.\,\ref{fig:vrad}, a large increase in radial velocity between
1992 and 1999 is obvious, followed by lower velocities after 2004. If
these changes are due to movement in a binary system, the data suggest
a long-period orbit.

In order to improve the time coverage, we searched for older published
radial velocities. A number of authors have measured the radial
velocity of $\theta^1$\,Ori\,C. These measurements have been
summarized by \citet{2002AstL...28..324V}. However, their Tab.\,1
contains errors in the mean time of the observations and partly
averages data obtained over several seasons.  The data of
\citet{1991ApJ...367..155A} have been obtained over many years and are
based only on the Balmer lines (which are in strong emission, and
highly variable).  The same is true for the data of
\citet{1926ApJ....64....1F}.  The average data reported by
\citet{2002AstL...28..324V} for both data sets are averaged over many
seasons and therefore not useful for our period study. Therefore, only
the data in Tab.~\,\ref{tab:vrad-old} remain. They are averages over
relatively short intervals and give reliable radial velocities.

\begin{table}
\caption{Published radial velocities. ``n'' is the number of spectra used.}
\begin{tabular}{llrr}
\hline
Reference & JD - 2\,400\,000 & rad.\ vel & n\\
\hline
\citet{1944ApJ....99...84S} & 30813 & 37.4 & 14 \\
\citet{1972ApJ...174L..79C} & 41302 & 26.4 & 10 \\
\citet{1991ApJS...75..965M} & 44932 & 9.2 & 6 \\
\hline
\hline
\end{tabular}
\label{tab:vrad-old}
\end{table}

A possible orbital origin for the trend in the radial velocity was
already discussed by \citet{2002MNRAS.333...55D} and
\citet{2007A&A...466..649K}, but we now cover a substantially longer
time interval. However, our data alone still do not allow us to determine an
unambiguous period.  An interferometric orbit was recently published by
\citet{2007A&A...466..649K}, and more recently revised with newer data
by \citet{2008ApJ...674L..97P}.  If we assume that the radial velocity
variations result from the orbit published by
\citet{2008ApJ...674L..97P}, we can fit an orbital solution.

\begin{figure}
\begin{center}
\includegraphics[height=8.8cm,angle=-90]{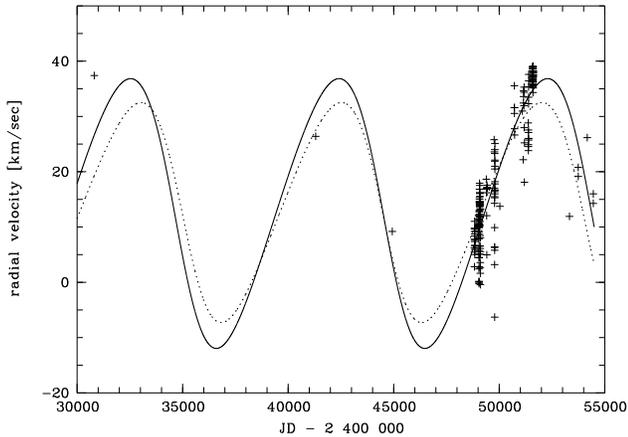}
\end{center}
\caption{Heliocentric radial velocity versus Julian date. The dashed line
  represents an orbital solution based on the parameters published by
  \citet{2008ApJ...674L..97P}, with only the systemic velocity
  $\gamma$ as a free parameter. The full line represents a solution
  which is within the errors compatible with interferometric orbit,
  but fits the radial velocities better.}
\label{fig:vrad-phase}
\end{figure}

In Fig.\,\ref{fig:vrad-phase}, the radial velocity (mean of all lines
in Table\,\ref{tab:vrad2} and the values from
Table\,\ref{tab:vrad-old}) is plotted versus Julian date. All
parameters derived from the interferometric orbit of
\citet{2008ApJ...674L..97P} have been kept fixed. We derived the $K$
value of the orbit from the relation $K = (2\pi/P) a \sin i /
\sqrt{1-e^2}$, using the parameters of \citet{2008ApJ...674L..97P} and
a distance of 440 pc \citep{2007MNRAS.376.1109J}. The only free
parameter was the systemic velocity $\gamma$ (dashed line).  It can be
seen that the radial velocity data are compatible with the
interferometric orbital parameters. If we optimize the radial-velocity
solution by varying all parameters, starting with the interferometric
parameters, we obtain another solution plotted in
Fig.\,\ref{fig:vrad-phase} (full line). Both solutions are summarized
in Table\,\ref{tab:orbits}.

Note that our solution is not unique. Within the errors given by
\citet{2008ApJ...674L..97P} several different radial velocity solutions
of similar quality are possible. Good solutions are also possible with
parameters which are incompatible with the interferometric orbit.
Therefore it is not possibly to give a reliable error estimate for our
solution. This is due to our very incomplete phase coverage and the
large short-term scatter. Both effects prevent a unique solution with
the available data. However, the period $P$ and the system velocity
$\gamma$ are relatively robust. 

Solutions with about half the period are also possible. In particular,
a highly eccentric orbit with a period near 11 years, similar to the
one proposed by \citet{2007A&A...466..649K}, fits the data reasonably
well, although not with the phases given by
\citet{2007A&A...466..649K} and with larger deviations than the
solution presented above. Such a short period seems to be excluded by
the interferometric data published by \citet{2008ApJ...674L..97P},
however.  Clearly, a longer time coverage is needed to derive a more reliable
solution.  

\begin{table*}
\caption{Summary of orbital parameters. Note that the radial velocity
  solution is not unique. The given solution is close to the
  parameters of \citet{2008ApJ...674L..97P}}
\begin{tabular}{lllllll}
\hline
Orbit & $P$ (days) & $T_0$ (JD) & $\omega$ (deg) & $\varepsilon$ & $K$ & $\gamma$ \\
\hline
\citet{2008ApJ...674L..97P} & 9497 $\pm$ 1461 & 2\,425\,610 $\pm$ 2154 & 96.9 $\pm$ 118.7 & 0.16 $\pm$ 0.14 & 19.9 & 13 $\pm$ 3 \\
radial velocity solution &  9880 & 2\,424\,932  & 99.3 & 0.142 & 24.4 & 13 \\
\hline
\end{tabular}
\label{tab:orbits}
\end{table*}

\section{Equivalent width variations of H$\alpha$ and He\,{\sc ii}$\lambda$4686}

\begin{figure}
\begin{center}
\includegraphics[height=8.8cm,angle=-90]{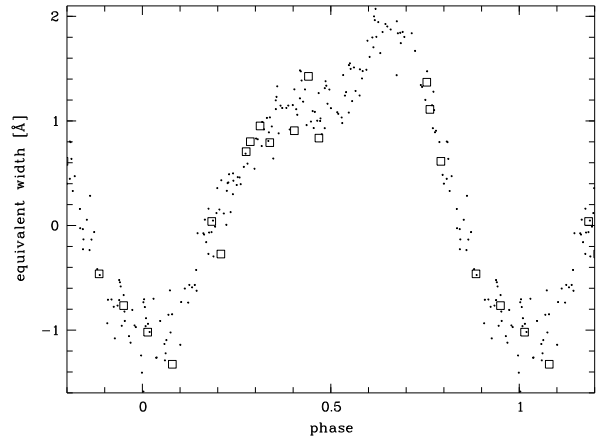}
\end{center}
\caption{The phase diagram of the equivalent width of
  H$\alpha$ as computed with a zero point of JD=2448833.0 and a
  period of 15.424 days. The new measurements are shown with the larger symbols.}
\label{fig:eqw}
\end{figure}

The equivalent width of the H$\alpha$ line of $\theta^1$\,Ori\,C shows
periodic variations, corresponding to the 15.4 day rotation period of
the star \citep{1993A&A...274L..29S}.  We measured the equivalent
width after subtraction of the nebular lines, following the procedure
described by \citet{1996A&A...312..539S}. The line was integrated
between 6545 and 6580 \AA\@.  The equivalent widths are listed in
Tab.\,\ref{tab:eqw2}. For completeness, we also include the measured
equivalent widths of He{\sc ii}\,$\lambda$4686 in the Table.  The
subtraction of the nebular lines is subjective and introduces an
error, which is difficult to quantify. The
error is small for the data with the highest resolution, but increases
with decreasing resolution. We combined our new results with the
published results to improve the period. From the AOV method
\citep{1989MNRAS.241..153S} we derive a best period of 15.424 $\pm$
0.001 days.  The phase diagram obtained with this period is shown in
Fig.\,\ref{fig:eqw}. The new period is, within the error bars, in
agreement with all published values, also with the value originally
obtained by \citet{1996A&A...312..539S} from IUE observations.

The error in the period is smaller than that obtained from previous
studies. However, because $\theta^1$\,Ori\,C is a member of a binary,
the times need to be corrected for light-travel effects due to
orbital motion. Because of the uncertain orbit, this correction has
not yet been applied.

As can be seen in Fig.\,\ref{fig:eqw}, the new measurements fit very
well in the phase diagram, although the scatter is larger than with
the higher resolution echelle data. This demonstrates that the
magneto-spheric emission of $\theta^1$\,Ori\,C as diagnosed by the
H$\alpha$ emission has been very stable over the past 15 years.

The published data of \citet{1972ApJ...174L..79C} are potentially
important for the period determination, because they extend the
covered time substantially. Their published line profiles of He\,{\sc
  ii}$\lambda$4686 (their Fig.\,2), show a blue-shifted emission
appearing between JD 2\,441\,284.91 and 2\,441\,287.88. According to
\citet{1996A&A...312..539S} (their Fig.\,7), this emission appears at
a phase of about 0.7. Together with our zero-point, this constrains the
period to 15.42 $< P <$ 15.426, in agreement with the period obtained
from H$\alpha$. Quantitative equivalent width measurements on the spectra of
\citet{1972ApJ...174L..79C} could provide stronger constraints.

\section{Other variations}

\begin{figure}
\begin{center}
\includegraphics[height=8.8cm,angle=-90]{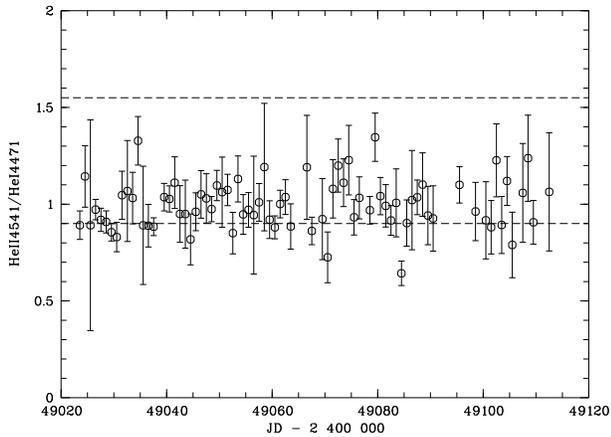}
\end{center}
\caption{The equivalent width ratio He\,{\sc ii}$\lambda$4541/He\,{\sc
    i}$\lambda$4471 versus time. The horizontal lines at 0.9 and 1.55 mark the
  approximate ratio for the spectral types O7V  and O5V, respectively. The measured ratio
  indicates a spectral type of about O7V. }
\label{fig:heiihei}
\end{figure}

The spectacular spectral type variations, occuring on a time-span of a
few days, reported by \citet{1981ApJ...243L..37W} are similar to the
variations reported by \citet{2003ApJ...588.1025W} for the other known
magnetic O star, HD\,191612. In the case of HD\,191612, the spectral
type variations are periodic with the rotational period.  For
  $\theta^1$\,Ori\,C, periodic equivalent variations of He{\sc i} and
  He{\sc ii} lines have also been reported
  \citep{1996A&A...312..539S}. However, the He\,{\sc i} and He\,{\sc
    ii} lines vary in phase and the ratio of these lines does not vary
  significantly with the rotational period
  \citep{1996A&A...312..539S}. This behavior has been explained by
  \citet{2006A&A...448..351S} by variable continuum emission from a
  disk. The variations found by \citet{1981ApJ...243L..37W} are
  different and thus probably had a different origin. In order to
  check our large data set for similar variations as reported by
  \citet{1981ApJ...243L..37W}, we searched for spectral type
  variations of $\theta^1$\,Ori\,C in our spectra. We analyzed the
  ratio He{\sc ii}\,$\lambda$4541/He{\sc i}\,$\lambda$4471. No
  variations similar to those observed by \citet{1981ApJ...243L..37W}
  were found. As an example, we show in Fig.\,\ref{fig:heiihei} the
  result of our measurements in the same time interval as the
  measurements in Fig.\,\ref{fig:para}. The measurements indicate a
  spectral type of about 07V, with little variation over time. For
  comparison, the ratios reported by \citet{1981ApJ...243L..37W} lie
  between 1.25 and 2.11, i.e.\@ they indicate a much earlier spectral
  type.  We have to conclude that such spectral-type variations are
  rare events in $\theta^1$\,Ori\,C.  

\section{Discussion and conclusions}

The long-term radial velocity variations of $\theta^1$\,Ori\,C are in
good agreement with the orbital motion expected from the published
interferometric orbit. Future radial velocity studies are very
important to complete the phase coverage.  Together with the
interferometric data \citep{2007A&A...466..649K,2008ApJ...674L..97P}
this should eventually provide data covering the full orbit of
$\theta^1$\,Ori\,C. The interferometric and the radial velocity data
are complementary, since some parameters are better constrained by
interferometry, while others are more sensitive to radial velocity
measurements.

We derive a system velocity of about 13 km s$^{-1}$ for
$\theta^1$\,Ori\,C, which is close to the velocity of the nebular
emission, but significantly below the radial velocity of the Orion
molecular cloud and the stars of the Orion nebula cluster, which have
a heliocentric radial velocity of about 30 km s$^{-1}$
\citep{2001ARA&A..39...99O}. While this discrepancy was already known
from previous measurements, our result indicates that the discrepancy
is not due to orbital motion. A large peculiar velocity of
$\theta^1$\,Ori\,C would have major effects on the ionization of the
Orion nebula \citep{2001ARA&A..39...99O}.  However, given the peculiar
spectrum of $\theta^1$\,Ori\,C, atmospheric effects can not be ruled
out completely. At least the occasional radial velocity deviations
towards smaller values (cf.\@ Fig.\,\ref{fig:para}) are probably due to
atmospheric distortions and could bias the measured systemic velocity
to slightly (by about 3 km s$^{-1}$) smaller values. In any case, the
good agreement of the interferometric orbit with the radial velocity
variations strongly indicates that the long-term radial-velocity
variations are due to orbital motion.

According to \citet{1997AJ....113.1733H}, $\theta^1$\,Ori\,C is younger
than 1 Myr. From its long period, we know that it is rotating very
slowly for an O-type star. A low rotation velocity of $v \sin $ = 24
km s$^{-1}$ has been found by direct spectroscopic analysis
\citep{2006A&A...448..351S}. Assuming that $\theta^1$\,Ori\,C was born
as a fast rotator, this suggests that strong magnetic braking must have
occurred.  

If it is assumed that the 538\,d spectral variability period of
HD\,191612 \citep{2007MNRAS.381..433H} is in fact the rotational period of that star (an
exceptionally long period for an O star), it would appear that magnetic fields are very
effective in slowing down the rotation rate. Interestingly, 
HD\,191612 also seems to be a member of a wide binary system
\citep{2007MNRAS.381..433H} with an orbital period of 1542\,d.

A factor of 10 decrease in rotational speed over its lifetime is
plausible for $\theta^1$\,Ori\,C.  If we assume that $p/\dot{p}$ was
constant with time, this leads to an $e$-folding time of of 500\,000
years or less.  On the other hand, a period change of 0.001 days in 20
years corresponds to a $p/\dot{p}$ of 300\,000 years. It seems
therefore possible to directly measure the deceleration age of
$\theta^1$\,Ori\,C in the foreseeable future. Equivalent width
measurements, especially at phase around 0.25 and 0.75, are needed.
At these phases the equivalent width changes rapidly with time,
therefore observations at these phases are particularly sensitive to
the exact value of the period.  We have demonstrated that observations
acquired using small telescopes are sufficient for this purpose.

However, in order to measure intrinsic period changes, the orbital
velocity of $\theta^1$\,Ori\,C (which introduces a variable Doppler
shift on the observed period) has to be determined with high
accuracy. An orbital velocity of only 3 km s$^{-1}$ already changes
the {\em observed\/} period by 10$^{-5}$ $P$, which is close to the
current accuracy of the period.  It is especially important to cover
the minimum of the radial velocity curve. Unpublished measurements
could be very valuable to fill the phase diagram.  If these
measurements are not available, a few more years of radial velocity
monitoring are needed.


\bibliographystyle{aa} \bibliography{thetaori}

\begin{acknowledgements}
This research has made use of the SIMBAD database,
operated at CDS, Strasbourg, France. GAW acknowledges
Discovery Grant from the Natural Science and Engineering Research Council of Canada.
\end{acknowledgements}

\begin{longtable}{lr@{.}lr@{.}lr@{.}lr@{.}lr@{.}lr@{.}lr@{.}lr@{.}ll}
\caption{Radial velocity measurements}\\
\hline
JD - 2\,400\,000  &  \multicolumn{4}{c}{C{\sc iv}$\lambda$5801} & \multicolumn{4}{c}{C{\sc iv}$\lambda$5812} & \multicolumn{4}{c}{He{\sc ii}$\lambda$5411}  & \multicolumn{4}{c}{O{\sc iii}$\lambda$5592}  &  instrument \\
\hline
\endfirsthead
\caption{continued}
\endhead
\endfoot
48822.934 & $13$ & $7$ & $\pm 1$ & $9$ & $3$ & $5$ & $\pm 2$ & $0$ & $0$ & $2$ & $ \pm3$ & $3$ & $13$ & $8$ & $\pm 2$ & $7$ & Flash \\
48823.915 & $7$ & $8$ & $\pm 1$ & $7$ & $2$ & $3$ & $\pm 1$ & $6$ & $3$ & $2$ & $ \pm2$ & $7$ & $11$ & $7$ & $\pm 2$ & $2$ & Flash \\
48824.919 & $2$ & $8$ & $\pm 1$ & $8$ & $1$ & $2$ & $\pm 1$ & $2$ & $3$ & $1$ & $ \pm2$ & $1$ & $4$ & $2$ & $\pm 2$ & $7$ & Flash \\
48825.912 & $3$ & $9$ & $\pm 1$ & $8$ & $3$ & $2$ & $\pm 1$ & $1$ & $7$ & $2$ & $ \pm1$ & $5$ & $7$ & $6$ & $\pm 1$ & $6$ & Flash \\
48829.915 & $11$ & $1$ & $\pm 2$ & $0$ & $7$ & $3$ & $\pm 1$ & $6$ & $10$ & $3$ & $ \pm2$ & $4$ & $8$ & $1$ & $\pm 2$ & $0$ & Flash \\
48830.915 & $10$ & $7$ & $\pm 1$ & $9$ & $2$ & $8$ & $\pm 0$ & $9$ & $16$ & $3$ & $ \pm1$ & $7$ & $9$ & $3$ & $\pm 2$ & $5$ & Flash \\
48833.903 & $10$ & $1$ & $\pm 1$ & $7$ & $8$ & $1$ & $\pm 0$ & $8$ & $12$ & $2$ & $ \pm1$ & $7$ & $13$ & $7$ & $\pm 1$ & $5$ & Flash \\
48835.909 & $9$ & $9$ & $\pm 0$ & $9$ & $5$ & $5$ & $\pm 1$ & $0$ & $14$ & $4$ & $ \pm1$ & $4$ & $8$ & $9$ & $\pm 1$ & $5$ & Flash \\
48836.902 & $9$ & $4$ & $\pm 0$ & $9$ & $6$ & $5$ & $\pm 2$ & $0$ & $10$ & $7$ & $ \pm1$ & $6$ & $6$ & $9$ & $\pm 1$ & $6$ & Flash \\
48837.903 & $9$ & $5$ & $\pm 1$ & $4$ & $9$ & $6$ & $\pm 1$ & $2$ & $6$ & $8$ & $ \pm1$ & $8$ & $9$ & $1$ & $\pm 2$ & $1$ & Flash \\
48838.892 & $9$ & $3$ & $\pm 0$ & $9$ & $4$ & $5$ & $\pm 2$ & $0$ & $11$ & $1$ & $ \pm1$ & $3$ & $13$ & $8$ & $\pm 1$ & $9$ & Flash \\
48839.896 & $7$ & $1$ & $\pm 1$ & $1$ & $2$ & $0$ & $\pm 0$ & $9$ & $11$ & $3$ & $ \pm1$ & $1$ & $9$ & $1$ & $\pm 1$ & $8$ & Flash \\
48841.889 & $5$ & $0$ & $\pm 0$ & $8$ & $6$ & $3$ & $\pm 2$ & $5$ & $4$ & $4$ & $ \pm1$ & $3$ & $4$ & $4$ & $\pm 2$ & $1$ & Flash \\
48844.888 & $5$ & $2$ & $\pm 1$ & $0$ & $3$ & $3$ & $\pm 1$ & $0$ & $8$ & $5$ & $ \pm1$ & $2$ & $6$ & $7$ & $\pm 1$ & $7$ & Flash \\
48845.878 & $6$ & $2$ & $\pm 0$ & $7$ & $5$ & $4$ & $\pm 0$ & $7$ & $9$ & $4$ & $ \pm1$ & $2$ & $5$ & $6$ & $\pm 1$ & $7$ & Flash \\
48847.876 & $4$ & $0$ & $\pm 0$ & $6$ & $3$ & $3$ & $\pm 0$ & $8$ & $14$ & $3$ & $ \pm1$ & $2$ & $8$ & $9$ & $\pm 1$ & $5$ & Flash \\
48848.879 & $5$ & $2$ & $\pm 0$ & $7$ & $3$ & $2$ & $\pm 1$ & $4$ & $9$ & $5$ & $ \pm1$ & $4$ & $8$ & $7$ & $\pm 1$ & $4$ & Flash \\
49023.551 & $13$ & $9$ & $\pm 0$ & $8$ & $9$ & $3$ & $\pm 2$ & $1$ & $16$ & $3$ & $ \pm1$ & $2$ & $9$ & $3$ & $\pm 1$ & $5$ & Flash \\
49024.535 & $10$ & $4$ & $\pm 0$ & $9$ & $14$ & $0$ & $\pm 1$ & $1$ & $10$ & $3$ & $ \pm1$ & $7$ & $10$ & $9$ & $\pm 2$ & $3$ & Flash \\
49025.551 & $9$ & $7$ & $\pm 0$ & $8$ & $9$ & $1$ & $\pm 2$ & $0$ & $6$ & $8$ & $ \pm1$ & $7$ & $6$ & $5$ & $\pm 1$ & $9$ & Flash \\
49026.535 & $8$ & $9$ & $\pm 0$ & $8$ & $7$ & $6$ & $\pm 1$ & $0$ & $10$ & $1$ & $ \pm1$ & $1$ & $10$ & $9$ & $\pm 1$ & $3$ & Flash \\
49027.543 & $10$ & $8$ & $\pm 0$ & $7$ & $6$ & $2$ & $\pm 1$ & $1$ & $13$ & $3$ & $ \pm1$ & $5$ & $9$ & $8$ & $\pm 1$ & $5$ & Flash \\
49028.543 & $11$ & $0$ & $\pm 0$ & $8$ & $7$ & $3$ & $\pm 0$ & $7$ & $8$ & $9$ & $ \pm1$ & $3$ & $7$ & $8$ & $\pm 1$ & $6$ & Flash \\
49029.535 & $9$ & $5$ & $\pm 1$ & $7$ & $7$ & $9$ & $\pm 2$ & $1$ & $12$ & $5$ & $ \pm1$ & $3$ & $6$ & $6$ & $\pm 1$ & $0$ & Flash \\
49030.543 & $12$ & $2$ & $\pm 0$ & $7$ & $7$ & $7$ & $\pm 0$ & $8$ & $15$ & $2$ & $ \pm1$ & $6$ & $5$ & $7$ & $\pm 1$ & $1$ & Flash \\
49031.539 & $11$ & $2$ & $\pm 0$ & $8$ & $8$ & $4$ & $\pm 0$ & $7$ & $13$ & $7$ & $ \pm1$ & $3$ & $8$ & $1$ & $\pm 0$ & $9$ & Flash \\
49032.562 & $8$ & $7$ & $\pm 0$ & $6$ & $7$ & $1$ & $\pm 0$ & $6$ & $14$ & $4$ & $ \pm1$ & $3$ & $6$ & $2$ & $\pm 2$ & $0$ & Flash \\
49033.555 & $9$ & $5$ & $\pm 1$ & $6$ & $6$ & $6$ & $\pm 1$ & $3$ & $15$ & $8$ & $ \pm1$ & $9$ & $11$ & $2$ & $\pm 1$ & $3$ & Flash \\
49034.578 & $9$ & $9$ & $\pm 0$ & $6$ & $9$ & $7$ & $\pm 0$ & $7$ & $16$ & $0$ & $ \pm1$ & $3$ & $10$ & $3$ & $\pm 2$ & $1$ & Flash \\
49035.547 & $9$ & $0$ & $\pm 0$ & $7$ & $10$ & $0$ & $\pm 0$ & $8$ & $12$ & $7$ & $ \pm1$ & $5$ & $10$ & $1$ & $\pm 1$ & $2$ & Flash \\
49036.543 & $8$ & $1$ & $\pm 0$ & $7$ & $6$ & $8$ & $\pm 1$ & $0$ & $6$ & $4$ & $ \pm1$ & $5$ & $10$ & $0$ & $\pm 0$ & $8$ & Flash \\
49037.535 & $7$ & $8$ & $\pm 0$ & $8$ & $4$ & $2$ & $\pm 1$ & $0$ & $3$ & $3$ & $ \pm1$ & $7$ & $7$ & $7$ & $\pm 1$ & $3$ & Flash \\
49038.609 & $8$ & $2$ & $\pm 1$ & $6$ & $-0$ & $9$ & $\pm 2$ & $0$ & $2$ & $4$ & $ \pm3$ & $7$ & $10$ & $7$ & $\pm 2$ & $1$ & Flash \\
49039.539 & $5$ & $7$ & $\pm 0$ & $7$ & $2$ & $8$ & $\pm 1$ & $0$ & $9$ & $9$ & $ \pm1$ & $2$ & $6$ & $8$ & $\pm 2$ & $2$ & Flash \\
49040.523 & $1$ & $8$ & $\pm 0$ & $8$ & $-2$ & $6$ & $\pm 1$ & $4$ & $5$ & $4$ & $ \pm1$ & $4$ & $5$ & $1$ & $\pm 1$ & $3$ & Flash \\
49041.527 & $4$ & $2$ & $\pm 0$ & $9$ & $-0$ & $4$ & $\pm 1$ & $0$ & $-5$ & $0$ & $ \pm1$ & $6$ & $1$ & $4$ & $\pm 1$ & $4$ & Flash \\
49042.527 & $0$ & $1$ & $\pm 1$ & $0$ & $-2$ & $5$ & $\pm 0$ & $7$ & $2$ & $5$ & $ \pm1$ & $7$ & $$ & $$ & $$ & $$ & Flash \\
49043.535 & $1$ & $2$ & $\pm 0$ & $7$ & $-3$ & $1$ & $\pm 0$ & $8$ & $5$ & $8$ & $ \pm1$ & $8$ & $-3$ & $2$ & $\pm 1$ & $3$ & Flash \\
49044.535 & $0$ & $7$ & $\pm 0$ & $7$ & $-3$ & $1$ & $\pm 0$ & $6$ & $2$ & $7$ & $ \pm1$ & $4$ & $-1$ & $2$ & $\pm 1$ & $1$ & Flash \\
49045.527 & $3$ & $0$ & $\pm 0$ & $6$ & $-0$ & $5$ & $\pm 0$ & $6$ & $8$ & $3$ & $ \pm1$ & $0$ & $0$ & $4$ & $\pm 1$ & $1$ & Flash \\
49046.523 & $3$ & $1$ & $\pm 0$ & $7$ & $2$ & $1$ & $\pm 1$ & $0$ & $9$ & $3$ & $ \pm1$ & $5$ & $3$ & $3$ & $\pm 1$ & $3$ & Flash \\
49047.527 & $3$ & $9$ & $\pm 0$ & $7$ & $0$ & $1$ & $\pm 0$ & $6$ & $11$ & $9$ & $ \pm1$ & $3$ & $2$ & $0$ & $\pm 1$ & $1$ & Flash \\
49048.523 & $4$ & $6$ & $\pm 0$ & $7$ & $4$ & $2$ & $\pm 0$ & $7$ & $12$ & $7$ & $ \pm1$ & $3$ & $4$ & $6$ & $\pm 0$ & $9$ & Flash \\
49049.520 & $7$ & $4$ & $\pm 0$ & $8$ & $5$ & $1$ & $\pm 0$ & $9$ & $14$ & $1$ & $ \pm1$ & $2$ & $7$ & $6$ & $\pm 1$ & $0$ & Flash \\
49050.523 & $10$ & $4$ & $\pm 0$ & $7$ & $8$ & $6$ & $\pm 0$ & $8$ & $14$ & $9$ & $ \pm1$ & $5$ & $11$ & $0$ & $\pm 1$ & $1$ & Flash \\
49051.547 & $11$ & $0$ & $\pm 0$ & $8$ & $8$ & $5$ & $\pm 0$ & $8$ & $10$ & $9$ & $ \pm1$ & $3$ & $9$ & $2$ & $\pm 1$ & $1$ & Flash \\
49052.547 & $11$ & $2$ & $\pm 1$ & $6$ & $7$ & $0$ & $\pm 0$ & $8$ & $14$ & $8$ & $ \pm1$ & $7$ & $9$ & $3$ & $\pm 1$ & $3$ & Flash \\
49053.523 & $12$ & $1$ & $\pm 0$ & $8$ & $7$ & $8$ & $\pm 1$ & $0$ & $13$ & $4$ & $ \pm2$ & $3$ & $8$ & $4$ & $\pm 1$ & $3$ & Flash \\
49054.512 & $13$ & $1$ & $\pm 1$ & $0$ & $7$ & $9$ & $\pm 0$ & $9$ & $11$ & $6$ & $ \pm1$ & $3$ & $15$ & $1$ & $\pm 1$ & $2$ & Flash \\
49055.516 & $11$ & $8$ & $\pm 1$ & $0$ & $10$ & $6$ & $\pm 1$ & $2$ & $12$ & $2$ & $ \pm1$ & $5$ & $7$ & $2$ & $\pm 1$ & $3$ & Flash \\
49056.512 & $11$ & $6$ & $\pm 1$ & $0$ & $8$ & $4$ & $\pm 1$ & $0$ & $12$ & $3$ & $ \pm1$ & $4$ & $7$ & $6$ & $\pm 1$ & $4$ & Flash \\
49057.520 & $10$ & $0$ & $\pm 1$ & $2$ & $7$ & $2$ & $\pm 0$ & $9$ & $11$ & $6$ & $ \pm1$ & $2$ & $12$ & $2$ & $\pm 1$ & $2$ & Flash \\
49058.512 & $12$ & $8$ & $\pm 0$ & $8$ & $8$ & $3$ & $\pm 1$ & $0$ & $11$ & $6$ & $ \pm1$ & $2$ & $15$ & $2$ & $\pm 2$ & $6$ & Flash \\
49059.512 & $11$ & $5$ & $\pm 0$ & $7$ & $10$ & $7$ & $\pm 0$ & $9$ & $12$ & $9$ & $ \pm1$ & $3$ & $15$ & $9$ & $\pm 1$ & $5$ & Flash \\
49060.516 & $14$ & $9$ & $\pm 0$ & $8$ & $10$ & $7$ & $\pm 0$ & $9$ & $16$ & $0$ & $ \pm2$ & $2$ & $10$ & $5$ & $\pm 2$ & $5$ & Flash \\
49061.512 & $16$ & $0$ & $\pm 0$ & $8$ & $5$ & $2$ & $\pm 1$ & $0$ & $19$ & $8$ & $ \pm1$ & $2$ & $10$ & $2$ & $\pm 1$ & $0$ & Flash \\
49062.512 & $13$ & $4$ & $\pm 0$ & $7$ & $11$ & $2$ & $\pm 0$ & $8$ & $19$ & $1$ & $ \pm1$ & $5$ & $14$ & $2$ & $\pm 1$ & $1$ & Flash \\
49063.516 & $13$ & $2$ & $\pm 0$ & $7$ & $12$ & $4$ & $\pm 0$ & $7$ & $19$ & $7$ & $ \pm1$ & $3$ & $13$ & $2$ & $\pm 1$ & $8$ & Flash \\
49065.512 & $16$ & $1$ & $\pm 0$ & $7$ & $14$ & $6$ & $\pm 0$ & $8$ & $19$ & $4$ & $ \pm1$ & $3$ & $17$ & $3$ & $\pm 1$ & $3$ & Flash \\
49066.578 & $15$ & $8$ & $\pm 1$ & $0$ & $12$ & $9$ & $\pm 1$ & $3$ & $16$ & $4$ & $ \pm2$ & $7$ & $16$ & $3$ & $\pm 2$ & $1$ & Flash \\
49067.512 & $14$ & $4$ & $\pm 0$ & $6$ & $13$ & $3$ & $\pm 0$ & $8$ & $17$ & $9$ & $ \pm1$ & $3$ & $14$ & $6$ & $\pm 1$ & $1$ & Flash \\
49068.512 & $15$ & $5$ & $\pm 0$ & $8$ & $12$ & $8$ & $\pm 1$ & $1$ & $15$ & $1$ & $ \pm1$ & $4$ & $14$ & $6$ & $\pm 1$ & $8$ & Flash \\
49069.516 & $16$ & $5$ & $\pm 1$ & $2$ & $12$ & $6$ & $\pm 1$ & $3$ & $23$ & $3$ & $ \pm4$ & $8$ & $19$ & $3$ & $\pm 1$ & $9$ & Flash \\
49070.516 & $15$ & $7$ & $\pm 1$ & $0$ & $12$ & $6$ & $\pm 0$ & $8$ & $15$ & $8$ & $ \pm1$ & $7$ & $12$ & $5$ & $\pm 1$ & $3$ & Flash \\
49071.516 & $13$ & $6$ & $\pm 1$ & $1$ & $9$ & $8$ & $\pm 0$ & $9$ & $17$ & $6$ & $ \pm1$ & $6$ & $14$ & $8$ & $\pm 1$ & $3$ & Flash \\
49072.516 & $12$ & $6$ & $\pm 1$ & $1$ & $10$ & $3$ & $\pm 1$ & $2$ & $17$ & $6$ & $ \pm1$ & $5$ & $16$ & $8$ & $\pm 3$ & $0$ & Flash \\
49073.527 & $14$ & $2$ & $\pm 0$ & $9$ & $12$ & $9$ & $\pm 0$ & $7$ & $15$ & $6$ & $ \pm1$ & $3$ & $9$ & $6$ & $\pm 1$ & $4$ & Flash \\
49074.520 & $14$ & $4$ & $\pm 0$ & $8$ & $13$ & $4$ & $\pm 0$ & $8$ & $15$ & $7$ & $ \pm1$ & $3$ & $11$ & $6$ & $\pm 2$ & $0$ & Flash \\
49075.520 & $15$ & $1$ & $\pm 0$ & $7$ & $8$ & $6$ & $\pm 1$ & $1$ & $17$ & $5$ & $ \pm1$ & $2$ & $8$ & $1$ & $\pm 4$ & $4$ & Flash \\
49076.516 & $15$ & $5$ & $\pm 0$ & $9$ & $11$ & $4$ & $\pm 0$ & $9$ & $15$ & $3$ & $ \pm1$ & $6$ & $11$ & $9$ & $\pm 1$ & $2$ & Flash \\
49078.520 & $15$ & $9$ & $\pm 0$ & $7$ & $13$ & $6$ & $\pm 0$ & $7$ & $21$ & $6$ & $ \pm1$ & $3$ & $13$ & $1$ & $\pm 1$ & $2$ & Flash \\
49079.516 & $15$ & $1$ & $\pm 0$ & $7$ & $12$ & $7$ & $\pm 1$ & $8$ & $22$ & $0$ & $ \pm1$ & $3$ & $14$ & $9$ & $\pm 1$ & $0$ & Flash \\
49080.512 & $15$ & $0$ & $\pm 0$ & $6$ & $13$ & $4$ & $\pm 0$ & $7$ & $21$ & $1$ & $ \pm1$ & $3$ & $15$ & $3$ & $\pm 1$ & $4$ & Flash \\
49081.520 & $16$ & $1$ & $\pm 0$ & $7$ & $14$ & $4$ & $\pm 0$ & $7$ & $22$ & $8$ & $ \pm1$ & $4$ & $16$ & $8$ & $\pm 1$ & $2$ & Flash \\
49082.516 & $15$ & $1$ & $\pm 0$ & $7$ & $14$ & $1$ & $\pm 1$ & $3$ & $16$ & $3$ & $ \pm1$ & $4$ & $12$ & $8$ & $\pm 1$ & $2$ & Flash \\
49083.508 & $12$ & $8$ & $\pm 1$ & $2$ & $14$ & $6$ & $\pm 1$ & $2$ & $17$ & $3$ & $ \pm1$ & $7$ & $13$ & $0$ & $\pm 1$ & $7$ & Flash \\
49084.504 & $19$ & $3$ & $\pm 1$ & $9$ & $12$ & $1$ & $\pm 0$ & $8$ & $9$ & $3$ & $ \pm1$ & $3$ & $14$ & $7$ & $\pm 1$ & $2$ & Flash \\
49085.508 & $19$ & $5$ & $\pm 1$ & $2$ & $11$ & $0$ & $\pm 1$ & $0$ & $14$ & $3$ & $ \pm1$ & $3$ & $17$ & $4$ & $\pm 1$ & $6$ & Flash \\
49086.504 & $11$ & $3$ & $\pm 1$ & $2$ & $11$ & $7$ & $\pm 1$ & $0$ & $9$ & $5$ & $ \pm1$ & $5$ & $15$ & $6$ & $\pm 1$ & $3$ & Flash \\
49087.500 & $11$ & $2$ & $\pm 1$ & $6$ & $8$ & $2$ & $\pm 0$ & $8$ & $5$ & $0$ & $ \pm1$ & $2$ & $16$ & $7$ & $\pm 2$ & $6$ & Flash \\
49088.508 & $11$ & $4$ & $\pm 1$ & $2$ & $7$ & $0$ & $\pm 0$ & $8$ & $14$ & $3$ & $ \pm1$ & $6$ & $14$ & $6$ & $\pm 1$ & $7$ & Flash \\
49089.520 & $13$ & $3$ & $\pm 1$ & $2$ & $9$ & $6$ & $\pm 0$ & $9$ & $13$ & $1$ & $ \pm1$ & $7$ & $10$ & $9$ & $\pm 2$ & $7$ & Flash \\
49090.504 & $13$ & $5$ & $\pm 1$ & $2$ & $9$ & $7$ & $\pm 0$ & $9$ & $11$ & $7$ & $ \pm1$ & $3$ & $9$ & $1$ & $\pm 4$ & $2$ & Flash \\
49095.504 & $14$ & $1$ & $\pm 0$ & $7$ & $11$ & $8$ & $\pm 1$ & $3$ & $14$ & $6$ & $ \pm1$ & $4$ & $17$ & $7$ & $\pm 2$ & $2$ & Flash \\
49098.539 & $7$ & $4$ & $\pm 1$ & $1$ & $6$ & $8$ & $\pm 1$ & $2$ & $11$ & $4$ & $ \pm2$ & $0$ & $8$ & $3$ & $\pm 1$ & $5$ & Flash \\
49100.504 & $4$ & $0$ & $\pm 1$ & $5$ & $3$ & $3$ & $\pm 1$ & $0$ & $11$ & $2$ & $ \pm1$ & $3$ & $7$ & $8$ & $\pm 1$ & $0$ & Flash \\
49101.504 & $6$ & $4$ & $\pm 1$ & $0$ & $3$ & $5$ & $\pm 1$ & $3$ & $7$ & $8$ & $ \pm1$ & $4$ & $6$ & $8$ & $\pm 1$ & $1$ & Flash \\
49102.496 & $2$ & $3$ & $\pm 1$ & $2$ & $0$ & $6$ & $\pm 0$ & $9$ & $7$ & $4$ & $ \pm1$ & $1$ & $4$ & $0$ & $\pm 1$ & $4$ & Flash \\
49103.496 & $0$ & $3$ & $\pm 0$ & $9$ & $1$ & $4$ & $\pm 0$ & $8$ & $7$ & $1$ & $ \pm1$ & $7$ & $1$ & $9$ & $\pm 1$ & $0$ & Flash \\
49104.508 & $0$ & $7$ & $\pm 1$ & $1$ & $-2$ & $0$ & $\pm 1$ & $0$ & $7$ & $7$ & $ \pm1$ & $1$ & $0$ & $1$ & $\pm 1$ & $1$ & Flash \\
49105.504 & $-2$ & $0$ & $\pm 2$ & $7$ & $-2$ & $0$ & $\pm 2$ & $8$ & $0$ & $0$ & $ \pm3$ & $8$ & $2$ & $3$ & $\pm 2$ & $2$ & Flash \\
49107.496 & $4$ & $0$ & $\pm 0$ & $6$ & $0$ & $8$ & $\pm 1$ & $1$ & $13$ & $7$ & $ \pm1$ & $6$ & $1$ & $8$ & $\pm 1$ & $6$ & Flash \\
49108.492 & $7$ & $7$ & $\pm 1$ & $4$ & $3$ & $6$ & $\pm 1$ & $3$ & $16$ & $2$ & $ \pm1$ & $7$ & $4$ & $8$ & $\pm 1$ & $4$ & Flash \\
49109.504 & $4$ & $2$ & $\pm 0$ & $9$ & $2$ & $3$ & $\pm 0$ & $8$ & $9$ & $2$ & $ \pm1$ & $5$ & $6$ & $5$ & $\pm 2$ & $3$ & Flash \\
49112.508 & $14$ & $1$ & $\pm 1$ & $1$ & $10$ & $0$ & $\pm 1$ & $2$ & $16$ & $4$ & $ \pm2$ & $0$ & $12$ & $4$ & $\pm 1$ & $4$ & Flash \\
49387.544 & $18$ & $3$ & $\pm 0$ & $8$ & $14$ & $2$ & $\pm 0$ & $7$ & $25$ & $7$ & $ \pm1$ & $2$ & $16$ & $4$ & $\pm 1$ & $2$ & Flash \\
49392.566 & $17$ & $2$ & $\pm 0$ & $8$ & $15$ & $9$ & $\pm 0$ & $9$ & $17$ & $8$ & $ \pm1$ & $5$ & $16$ & $5$ & $\pm 1$ & $5$ & Flash \\
49395.539 & $14$ & $8$ & $\pm 2$ & $7$ & $13$ & $8$ & $\pm 3$ & $0$ & $20$ & $3$ & $ \pm1$ & $8$ & $15$ & $9$ & $\pm 2$ & $4$ & Flash \\
49401.526 & $16$ & $9$ & $\pm 0$ & $7$ & $15$ & $4$ & $\pm 0$ & $9$ & $22$ & $1$ & $ \pm1$ & $2$ & $13$ & $7$ & $\pm 1$ & $3$ & Flash \\
49406.519 & $11$ & $9$ & $\pm 0$ & $7$ & $10$ & $1$ & $\pm 0$ & $8$ & $13$ & $6$ & $ \pm1$ & $5$ & $12$ & $6$ & $\pm 1$ & $0$ & Flash \\
49413.517 & $5$ & $0$ & $\pm 0$ & $9$ & $0$ & $4$ & $\pm 0$ & $8$ & $7$ & $3$ & $ \pm1$ & $4$ & $7$ & $2$ & $\pm 1$ & $2$ & Flash \\
49422.549 & $17$ & $3$ & $\pm 0$ & $8$ & $15$ & $9$ & $\pm 0$ & $8$ & $19$ & $0$ & $ \pm1$ & $5$ & $18$ & $0$ & $\pm 1$ & $8$ & Flash \\
49429.517 & $16$ & $5$ & $\pm 0$ & $7$ & $15$ & $3$ & $\pm 1$ & $2$ & $18$ & $2$ & $ \pm1$ & $3$ & $18$ & $8$ & $\pm 9$ & $2$ & Flash \\
49433.520 & $19$ & $1$ & $\pm 0$ & $7$ & $11$ & $5$ & $\pm 1$ & $8$ & $22$ & $2$ & $ \pm1$ & $1$ & $15$ & $4$ & $\pm 1$ & $4$ & Flash \\
49759.567 & $$ & $$ & $$ & $$ & $$ & $$ & $$ & $$ & $25$ & $8$ & $ \pm1$ & $8$ & $$ & $$ & $$ & $$ & Heros \\
49771.591 & $$ & $$ & $$ & $$ & $$ & $$ & $$ & $$ & $23$ & $6$ & $ \pm2$ & $1$ & $$ & $$ & $$ & $$ & Heros \\
49776.570 & $$ & $$ & $$ & $$ & $$ & $$ & $$ & $$ & $17$ & $0$ & $ \pm1$ & $7$ & $$ & $$ & $$ & $$ & Heros \\
49777.574 & $$ & $$ & $$ & $$ & $$ & $$ & $$ & $$ & $14$ & $3$ & $ \pm1$ & $7$ & $$ & $$ & $$ & $$ & Heros \\
49778.519 & $$ & $$ & $$ & $$ & $$ & $$ & $$ & $$ & $15$ & $5$ & $ \pm1$ & $5$ & $$ & $$ & $$ & $$ & Heros \\
49779.518 & $$ & $$ & $$ & $$ & $$ & $$ & $$ & $$ & $6$ & $3$ & $ \pm2$ & $0$ & $$ & $$ & $$ & $$ & Heros \\
49780.518 & $$ & $$ & $$ & $$ & $$ & $$ & $$ & $$ & $9$ & $9$ & $ \pm1$ & $5$ & $$ & $$ & $$ & $$ & Heros \\
49781.517 & $$ & $$ & $$ & $$ & $$ & $$ & $$ & $$ & $3$ & $2$ & $ \pm2$ & $3$ & $$ & $$ & $$ & $$ & Heros \\
49782.520 & $$ & $$ & $$ & $$ & $$ & $$ & $$ & $$ & $6$ & $4$ & $ \pm1$ & $6$ & $$ & $$ & $$ & $$ & Heros \\
49783.517 & $$ & $$ & $$ & $$ & $$ & $$ & $$ & $$ & $5$ & $8$ & $ \pm2$ & $2$ & $$ & $$ & $$ & $$ & Heros \\
49784.517 & $$ & $$ & $$ & $$ & $$ & $$ & $$ & $$ & $-6$ & $3$ & $ \pm2$ & $9$ & $$ & $$ & $$ & $$ & Heros \\
49785.517 & $$ & $$ & $$ & $$ & $$ & $$ & $$ & $$ & $16$ & $4$ & $ \pm2$ & $1$ & $$ & $$ & $$ & $$ & Heros \\
49786.520 & $$ & $$ & $$ & $$ & $$ & $$ & $$ & $$ & $15$ & $3$ & $ \pm2$ & $0$ & $$ & $$ & $$ & $$ & Heros \\
49787.543 & $$ & $$ & $$ & $$ & $$ & $$ & $$ & $$ & $23$ & $3$ & $ \pm2$ & $1$ & $$ & $$ & $$ & $$ & Heros \\
49788.522 & $$ & $$ & $$ & $$ & $$ & $$ & $$ & $$ & $16$ & $6$ & $ \pm1$ & $9$ & $$ & $$ & $$ & $$ & Heros \\
49789.519 & $$ & $$ & $$ & $$ & $$ & $$ & $$ & $$ & $16$ & $0$ & $ \pm1$ & $8$ & $$ & $$ & $$ & $$ & Heros \\
49790.505 & $$ & $$ & $$ & $$ & $$ & $$ & $$ & $$ & $24$ & $0$ & $ \pm2$ & $0$ & $$ & $$ & $$ & $$ & Heros \\
49791.545 & $$ & $$ & $$ & $$ & $$ & $$ & $$ & $$ & $21$ & $9$ & $ \pm2$ & $5$ & $$ & $$ & $$ & $$ & Heros \\
49792.518 & $$ & $$ & $$ & $$ & $$ & $$ & $$ & $$ & $21$ & $0$ & $ \pm2$ & $1$ & $$ & $$ & $$ & $$ & Heros \\
49793.510 & $$ & $$ & $$ & $$ & $$ & $$ & $$ & $$ & $20$ & $5$ & $ \pm3$ & $8$ & $$ & $$ & $$ & $$ & Heros \\
49794.506 & $$ & $$ & $$ & $$ & $$ & $$ & $$ & $$ & $22$ & $1$ & $ \pm1$ & $5$ & $$ & $$ & $$ & $$ & Heros \\
49795.509 & $$ & $$ & $$ & $$ & $$ & $$ & $$ & $$ & $18$ & $2$ & $ \pm5$ & $6$ & $$ & $$ & $$ & $$ & Heros \\
49796.503 & $$ & $$ & $$ & $$ & $$ & $$ & $$ & $$ & $25$ & $1$ & $ \pm1$ & $7$ & $$ & $$ & $$ & $$ & Heros \\
49797.506 & $$ & $$ & $$ & $$ & $$ & $$ & $$ & $$ & $17$ & $8$ & $ \pm1$ & $7$ & $$ & $$ & $$ & $$ & Heros \\
49798.499 & $$ & $$ & $$ & $$ & $$ & $$ & $$ & $$ & $16$ & $7$ & $ \pm1$ & $7$ & $$ & $$ & $$ & $$ & Heros \\
50030.595 & $14$ & $4$ & $\pm 0$ & $3$ & $11$ & $2$ & $\pm 0$ & $3$ & $16$ & $2$ & $ \pm0$ & $7$ & $13$ & $2$ & $\pm 0$ & $5$ & Elodie \\
50716.651 & $$ & $$ & $$ & $$ & $$ & $$ & $$ & $$ & $43$ & $6$ & $ \pm2$ & $2$ & $27$ & $5$ & $\pm 2$ & $0$ & Heros \\
50717.618 & $$ & $$ & $$ & $$ & $$ & $$ & $$ & $$ & $33$ & $9$ & $ \pm1$ & $7$ & $29$ & $3$ & $\pm 2$ & $3$ & Heros \\
50718.590 & $$ & $$ & $$ & $$ & $$ & $$ & $$ & $$ & $30$ & $9$ & $ \pm1$ & $9$ & $32$ & $3$ & $\pm 1$ & $7$ & Heros \\
50719.612 & $$ & $$ & $$ & $$ & $$ & $$ & $$ & $$ & $30$ & $2$ & $ \pm1$ & $7$ & $30$ & $9$ & $\pm 2$ & $0$ & Heros \\
50726.583 & $31$ & $7$ & $\pm 1$ & $6$ & $23$ & $9$ & $\pm 1$ & $2$ & $30$ & $7$ & $ \pm2$ & $4$ & $24$ & $8$ & $\pm 2$ & $4$ & Heros \\
50727.597 & $27$ & $2$ & $\pm 1$ & $0$ & $26$ & $1$ & $\pm 1$ & $2$ & $$ & $$ & $$ & $$ & $$ & $$ & $$ & $$ & Heros \\
51097.892 & $34$ & $7$ & $\pm 0$ & $3$ & $32$ & $2$ & $\pm 0$ & $3$ & $37$ & $9$ & $ \pm0$ & $5$ & $19$ & $1$ & $\pm 3$ & $7$ & Feros \\
51134.880 & $23$ & $3$ & $\pm 0$ & $2$ & $20$ & $7$ & $\pm 0$ & $3$ & $24$ & $1$ & $ \pm0$ & $5$ & $20$ & $6$ & $\pm 0$ & $5$ & Feros \\
51145.866 & $32$ & $1$ & $\pm 0$ & $3$ & $28$ & $7$ & $\pm 0$ & $2$ & $38$ & $3$ & $ \pm0$ & $5$ & $30$ & $6$ & $\pm 0$ & $5$ & Feros \\
51148.801 & $35$ & $1$ & $\pm 0$ & $3$ & $32$ & $0$ & $\pm 0$ & $2$ & $38$ & $1$ & $ \pm0$ & $5$ & $33$ & $3$ & $\pm 0$ & $5$ & Feros \\
51151.776 & $34$ & $7$ & $\pm 0$ & $4$ & $31$ & $8$ & $\pm 0$ & $3$ & $34$ & $3$ & $ \pm0$ & $6$ & $33$ & $6$ & $\pm 0$ & $6$ & Feros \\
51172.579 & $34$ & $4$ & $\pm 0$ & $5$ & $28$ & $1$ & $\pm 0$ & $6$ & $34$ & $6$ & $ \pm0$ & $7$ & $30$ & $6$ & $\pm 1$ & $1$ & Feros \\
51177.579 & $34$ & $4$ & $\pm 0$ & $3$ & $32$ & $2$ & $\pm 0$ & $3$ & $40$ & $5$ & $ \pm0$ & $7$ & $34$ & $1$ & $\pm 0$ & $6$ & Feros \\
51179.593 & $34$ & $0$ & $\pm 0$ & $3$ & $31$ & $3$ & $\pm 0$ & $3$ & $38$ & $4$ & $ \pm0$ & $6$ & $33$ & $8$ & $\pm 0$ & $6$ & Feros \\
51180.513 & $33$ & $5$ & $\pm 0$ & $4$ & $31$ & $4$ & $\pm 0$ & $3$ & $35$ & $5$ & $ \pm0$ & $8$ & $11$ & $5$ & $\pm 6$ & $5$ & Feros \\
51191.700 & $26$ & $0$ & $\pm 0$ & $2$ & $22$ & $4$ & $\pm 0$ & $2$ & $0$ & $5$ & $ \pm0$ & $5$ & $23$ & $5$ & $\pm 0$ & $5$ & Feros \\
51192.551 & $24$ & $5$ & $\pm 0$ & $2$ & $22$ & $2$ & $\pm 0$ & $2$ & $31$ & $5$ & $ \pm0$ & $5$ & $22$ & $7$ & $\pm 0$ & $6$ & Feros \\
51373.931 & $32$ & $5$ & $\pm 0$ & $4$ & $29$ & $3$ & $\pm 0$ & $4$ & $35$ & $6$ & $ \pm0$ & $6$ & $17$ & $9$ & $\pm 2$ & $2$ & Feros \\
51380.936 & $25$ & $2$ & $\pm 0$ & $3$ & $24$ & $5$ & $\pm 0$ & $3$ & $27$ & $8$ & $ \pm0$ & $6$ & $22$ & $1$ & $\pm 0$ & $7$ & Feros \\
51381.937 & $25$ & $9$ & $\pm 0$ & $3$ & $23$ & $6$ & $\pm 0$ & $3$ & $27$ & $4$ & $ \pm0$ & $6$ & $18$ & $4$ & $\pm 1$ & $5$ & Feros \\
51382.939 & $26$ & $1$ & $\pm 0$ & $4$ & $23$ & $2$ & $\pm 0$ & $4$ & $26$ & $0$ & $ \pm0$ & $6$ & $22$ & $9$ & $\pm 0$ & $9$ & Feros \\
51383.932 & $28$ & $4$ & $\pm 0$ & $4$ & $24$ & $9$ & $\pm 0$ & $4$ & $25$ & $8$ & $ \pm0$ & $5$ & $24$ & $9$ & $\pm 0$ & $6$ & Feros \\
51384.928 & $28$ & $0$ & $\pm 0$ & $4$ & $24$ & $3$ & $\pm 0$ & $4$ & $26$ & $0$ & $ \pm0$ & $5$ & $25$ & $8$ & $\pm 0$ & $7$ & Feros \\
51385.935 & $28$ & $3$ & $\pm 0$ & $4$ & $26$ & $2$ & $\pm 0$ & $3$ & $29$ & $1$ & $ \pm0$ & $6$ & $24$ & $8$ & $\pm 0$ & $9$ & Feros \\
51386.946 & $29$ & $2$ & $\pm 0$ & $3$ & $27$ & $2$ & $\pm 0$ & $3$ & $30$ & $2$ & $ \pm0$ & $5$ & $23$ & $6$ & $\pm 0$ & $8$ & Feros \\
51389.933 & $33$ & $7$ & $\pm 0$ & $3$ & $30$ & $4$ & $\pm 0$ & $3$ & $36$ & $4$ & $ \pm0$ & $5$ & $28$ & $7$ & $\pm 0$ & $4$ & Feros \\
51390.954 & $34$ & $8$ & $\pm 0$ & $3$ & $31$ & $6$ & $\pm 0$ & $2$ & $38$ & $7$ & $ \pm0$ & $6$ & $28$ & $8$ & $\pm 0$ & $5$ & Feros \\
51391.947 & $35$ & $6$ & $\pm 0$ & $3$ & $33$ & $0$ & $\pm 0$ & $3$ & $41$ & $1$ & $ \pm0$ & $5$ & $31$ & $7$ & $\pm 0$ & $5$ & Feros \\
51392.900 & $35$ & $3$ & $\pm 0$ & $3$ & $33$ & $4$ & $\pm 0$ & $3$ & $2$ & $1$ & $ \pm0$ & $5$ & $32$ & $1$ & $\pm 0$ & $5$ & Feros \\
51393.950 & $37$ & $5$ & $\pm 0$ & $3$ & $34$ & $9$ & $\pm 0$ & $2$ & $43$ & $5$ & $ \pm0$ & $5$ & $29$ & $8$ & $\pm 1$ & $3$ & Feros \\
51394.935 & $37$ & $9$ & $\pm 0$ & $3$ & $35$ & $2$ & $\pm 0$ & $3$ & $43$ & $1$ & $ \pm0$ & $5$ & $34$ & $2$ & $\pm 0$ & $7$ & Feros \\
53329.563 & $11$ & $6$ & $\pm 0$ & $8$ & $9$ & $3$ & $\pm 1$ & $4$ & $15$ & $6$ & $ \pm0$ & $9$ & $11$ & $2$ & $\pm 0$ & $7$ & Elodie \\
53740.534 & $21$ & $6$ & $\pm 0$ & $2$ & $18$ & $5$ & $\pm 0$ & $4$ & $21$ & $6$ & $ \pm0$ & $9$ & $21$ & $4$ & $\pm 0$ & $6$ & Feros \\
51578.422 & $37$ & $8$ & $\pm 0$ & $7$ & $35$ & $3$ & $\pm 0$ & $5$ & $39$ & $3$ & $ \pm1$ & $1$ & $36$ & $0$ & $\pm 0$ & $9$ & Musicos \\
51578.454 & $36$ & $9$ & $\pm 0$ & $7$ & $35$ & $2$ & $\pm 0$ & $6$ & $37$ & $3$ & $ \pm1$ & $2$ & $37$ & $5$ & $\pm 0$ & $7$ & Musicos \\
51579.393 & $40$ & $1$ & $\pm 0$ & $8$ & $37$ & $9$ & $\pm 0$ & $5$ & $38$ & $2$ & $ \pm1$ & $1$ & $38$ & $5$ & $\pm 0$ & $7$ & Musicos \\
51579.423 & $39$ & $9$ & $\pm 0$ & $6$ & $36$ & $6$ & $\pm 0$ & $5$ & $40$ & $1$ & $ \pm1$ & $1$ & $39$ & $1$ & $\pm 0$ & $7$ & Musicos \\
51579.454 & $38$ & $4$ & $\pm 0$ & $6$ & $36$ & $9$ & $\pm 0$ & $5$ & $40$ & $3$ & $ \pm1$ & $1$ & $37$ & $5$ & $\pm 0$ & $6$ & Musicos \\
51587.374 & $37$ & $6$ & $\pm 0$ & $7$ & $35$ & $9$ & $\pm 0$ & $6$ & $34$ & $2$ & $ \pm1$ & $2$ & $38$ & $7$ & $\pm 0$ & $9$ & Musicos \\
51590.382 & $39$ & $4$ & $\pm 0$ & $7$ & $35$ & $8$ & $\pm 2$ & $6$ & $33$ & $3$ & $ \pm1$ & $1$ & $37$ & $4$ & $\pm 0$ & $9$ & Musicos \\
51590.413 & $39$ & $0$ & $\pm 0$ & $7$ & $35$ & $5$ & $\pm 7$ & $7$ & $33$ & $0$ & $ \pm1$ & $2$ & $37$ & $5$ & $\pm 0$ & $9$ & Musicos \\
51596.391 & $41$ & $0$ & $\pm 0$ & $6$ & $37$ & $2$ & $\pm 0$ & $6$ & $37$ & $0$ & $ \pm1$ & $2$ & $39$ & $2$ & $\pm 0$ & $6$ & Musicos \\
51596.421 & $40$ & $2$ & $\pm 0$ & $6$ & $37$ & $6$ & $\pm 0$ & $5$ & $38$ & $3$ & $ \pm1$ & $1$ & $40$ & $2$ & $\pm 0$ & $6$ & Musicos \\
51596.452 & $40$ & $3$ & $\pm 0$ & $6$ & $37$ & $6$ & $\pm 0$ & $6$ & $37$ & $9$ & $ \pm1$ & $2$ & $40$ & $3$ & $\pm 0$ & $6$ & Musicos \\
51597.351 & $41$ & $2$ & $\pm 0$ & $6$ & $38$ & $0$ & $\pm 0$ & $5$ & $36$ & $6$ & $ \pm1$ & $3$ & $38$ & $8$ & $\pm 0$ & $6$ & Musicos \\
51597.382 & $41$ & $9$ & $\pm 0$ & $6$ & $37$ & $9$ & $\pm 0$ & $5$ & $36$ & $4$ & $ \pm1$ & $2$ & $39$ & $6$ & $\pm 0$ & $7$ & Musicos \\
51597.413 & $41$ & $3$ & $\pm 0$ & $7$ & $38$ & $3$ & $\pm 0$ & $5$ & $35$ & $5$ & $ \pm1$ & $2$ & $40$ & $8$ & $\pm 0$ & $7$ & Musicos \\
51597.444 & $41$ & $9$ & $\pm 0$ & $7$ & $37$ & $7$ & $\pm 0$ & $5$ & $35$ & $1$ & $ \pm1$ & $3$ & $39$ & $9$ & $\pm 0$ & $7$ & Musicos \\
51599.353 & $41$ & $4$ & $\pm 0$ & $8$ & $35$ & $4$ & $\pm 0$ & $7$ & $32$ & $0$ & $ \pm1$ & $2$ & $38$ & $7$ & $\pm 0$ & $6$ & Musicos \\
51599.384 & $41$ & $2$ & $\pm 0$ & $7$ & $35$ & $3$ & $\pm 0$ & $7$ & $31$ & $0$ & $ \pm1$ & $4$ & $38$ & $6$ & $\pm 1$ & $0$ & Musicos \\
51599.416 & $40$ & $6$ & $\pm 0$ & $7$ & $34$ & $6$ & $\pm 0$ & $7$ & $27$ & $7$ & $ \pm1$ & $4$ & $39$ & $3$ & $\pm 1$ & $1$ & Musicos \\
51600.346 & $40$ & $0$ & $\pm 0$ & $8$ & $35$ & $7$ & $\pm 2$ & $8$ & $32$ & $2$ & $ \pm1$ & $1$ & $39$ & $6$ & $\pm 1$ & $2$ & Musicos \\
51600.378 & $40$ & $2$ & $\pm 0$ & $8$ & $36$ & $7$ & $\pm 0$ & $9$ & $33$ & $7$ & $ \pm1$ & $2$ & $38$ & $5$ & $\pm 1$ & $3$ & Musicos \\
51600.410 & $38$ & $7$ & $\pm 0$ & $8$ & $36$ & $1$ & $\pm 5$ & $6$ & $31$ & $0$ & $ \pm1$ & $2$ & $37$ & $5$ & $\pm 0$ & $9$ & Musicos \\
51601.361 & $38$ & $8$ & $\pm 0$ & $8$ & $34$ & $6$ & $\pm 0$ & $8$ & $30$ & $6$ & $ \pm1$ & $2$ & $38$ & $8$ & $\pm 1$ & $2$ & Musicos \\
51601.392 & $39$ & $3$ & $\pm 0$ & $8$ & $34$ & $8$ & $\pm 0$ & $8$ & $31$ & $8$ & $ \pm1$ & $2$ & $39$ & $1$ & $\pm 1$ & $3$ & Musicos \\
51601.423 & $38$ & $3$ & $\pm 0$ & $7$ & $33$ & $4$ & $\pm 1$ & $1$ & $29$ & $0$ & $ \pm1$ & $2$ & $36$ & $6$ & $\pm 1$ & $2$ & Musicos \\
51602.349 & $37$ & $8$ & $\pm 0$ & $7$ & $34$ & $4$ & $\pm 0$ & $6$ & $31$ & $6$ & $ \pm1$ & $1$ & $38$ & $9$ & $\pm 1$ & $2$ & Musicos \\
51602.388 & $38$ & $4$ & $\pm 0$ & $6$ & $34$ & $6$ & $\pm 0$ & $5$ & $31$ & $0$ & $ \pm1$ & $1$ & $37$ & $2$ & $\pm 0$ & $9$ & Musicos \\
51602.419 & $38$ & $2$ & $\pm 0$ & $6$ & $35$ & $0$ & $\pm 0$ & $7$ & $30$ & $9$ & $ \pm1$ & $2$ & $39$ & $1$ & $\pm 1$ & $1$ & Musicos \\
51603.377 & $38$ & $3$ & $\pm 0$ & $7$ & $35$ & $7$ & $\pm 4$ & $0$ & $31$ & $7$ & $ \pm1$ & $3$ & $39$ & $1$ & $\pm 1$ & $1$ & Musicos \\
51603.408 & $38$ & $9$ & $\pm 0$ & $7$ & $36$ & $2$ & $\pm 2$ & $9$ & $34$ & $1$ & $ \pm1$ & $2$ & $39$ & $5$ & $\pm 1$ & $2$ & Musicos \\
51603.439 & $38$ & $2$ & $\pm 0$ & $8$ & $34$ & $9$ & $\pm 6$ & $2$ & $29$ & $2$ & $ \pm1$ & $4$ & $38$ & $3$ & $\pm 1$ & $2$ & Musicos \\
51606.353 & $38$ & $6$ & $\pm 0$ & $7$ & $36$ & $5$ & $\pm 5$ & $2$ & $37$ & $4$ & $ \pm1$ & $1$ & $36$ & $4$ & $\pm 0$ & $8$ & Musicos \\
51606.384 & $40$ & $9$ & $\pm 0$ & $7$ & $37$ & $4$ & $\pm 10$ & $2$ & $38$ & $1$ & $ \pm1$ & $0$ & $36$ & $5$ & $\pm 0$ & $8$ & Musicos \\
51606.416 & $39$ & $7$ & $\pm 0$ & $7$ & $36$ & $2$ & $\pm 3$ & $7$ & $37$ & $2$ & $ \pm1$ & $1$ & $37$ & $2$ & $\pm 1$ & $0$ & Musicos \\
51608.370 & $37$ & $4$ & $\pm 0$ & $6$ & $35$ & $7$ & $\pm 0$ & $5$ & $40$ & $0$ & $ \pm1$ & $1$ & $36$ & $7$ & $\pm 0$ & $4$ & Musicos \\
51608.401 & $38$ & $3$ & $\pm 0$ & $7$ & $36$ & $0$ & $\pm 0$ & $5$ & $40$ & $5$ & $ \pm1$ & $1$ & $37$ & $0$ & $\pm 0$ & $5$ & Musicos \\
51608.431 & $38$ & $3$ & $\pm 0$ & $6$ & $36$ & $3$ & $\pm 0$ & $6$ & $39$ & $0$ & $ \pm1$ & $2$ & $36$ & $3$ & $\pm 0$ & $6$ & Musicos \\
51609.361 & $39$ & $3$ & $\pm 0$ & $6$ & $36$ & $2$ & $\pm 0$ & $6$ & $38$ & $3$ & $ \pm1$ & $7$ & $39$ & $1$ & $\pm 0$ & $6$ & Musicos \\
51609.391 & $38$ & $4$ & $\pm 0$ & $6$ & $36$ & $4$ & $\pm 0$ & $5$ & $39$ & $7$ & $ \pm1$ & $1$ & $37$ & $5$ & $\pm 0$ & $6$ & Musicos \\
51609.422 & $39$ & $7$ & $\pm 0$ & $7$ & $36$ & $4$ & $\pm 0$ & $7$ & $36$ & $5$ & $ \pm1$ & $2$ & $39$ & $0$ & $\pm 0$ & $6$ & Musicos \\
53744.891 & $20$ & $1$ & $\pm 0$ & $2$ & $17$ & $0$ & $\pm 0$ & $2$ & $21$ & $3$ & $ \pm0$ & $5$ & $18$ & $2$ & $\pm 0$ & $4$ & Espadons \\
54169.854 & $26$ & $0$ & $\pm 0$ & $2$ & $23$ & $1$ & $\pm 0$ & $2$ & $32$ & $2$ & $ \pm0$ & $5$ & $23$ & $4$ & $\pm 0$ & $4$ & Espadons \\
54456.748 & $15$ & $4$ & $\pm 0$ & $2$ & $11$ & $9$ & $\pm 0$ & $2$ & $16$ & $9$ & $ \pm0$ & $4$ & $12$ & $9$ & $\pm 0$ & $3$ & Espadons \\
54457.748 & $17$ & $1$ & $\pm 0$ & $2$ & $13$ & $8$ & $\pm 0$ & $2$ & $19$ & $5$ & $ \pm0$ & $5$ & $13$ & $6$ & $\pm 0$ & $2$ & Espadons \\
\hline
\hline
\label{tab:vrad2}
\end{longtable}

\begin{longtable}{lr@{.}lr@{.}ll}
\caption{Equivalent width measurements}\\
\hline
JD $-$ 2\,400\,000  &  \multicolumn{2}{c}{H$\alpha$} &  \multicolumn{2}{c}{He{\sc ii}$\lambda$4686}  &  instrument \\
\hline
\endfirsthead
\caption{continued}
\endhead
\endfoot
48822.934 & $0$ & $640$ & $0$ & $309$ & Flash \\
48823.915 & $1$ & $059$ & $0$ & $202$ & Flash \\
48824.919 & $0$ & $873$ & $0$ & $195$ & Flash \\
48825.912 & $1$ & $276$ & $0$ & $361$ & Flash \\
48829.915 & $0$ & $400$ & $0$ & $131$ & Flash \\
48830.915 & $-0$ & $134$ & $0$ & $033$ & Flash \\
48833.903 & $-1$ & $133$ & $-0$ & $140$ & Flash \\
48835.909 & $-0$ & $296$ & $-0$ & $032$ & Flash \\
48836.902 & $0$ & $462$ & $0$ & $136$ & Flash \\
48837.903 & $0$ & $990$ & $0$ & $258$ & Flash \\
48838.892 & $1$ & $124$ & $0$ & $238$ & Flash \\
48839.896 & $0$ & $994$ & $0$ & $258$ & Flash \\
48841.889 & $1$ & $243$ & $0$ & $247$ & Flash \\
48844.888 & $0$ & $931$ & $0$ & $231$ & Flash \\
48845.878 & $-0$ & $026$ & $0$ & $182$ & Flash \\
48847.876 & $-1$ & $057$ & $0$ & $016$ & Flash \\
48848.879 & $-0$ & $700$ & $0$ & $196$ & Flash \\
49023.551 & $1$ & $113$ & $0$ & $214$ & Flash \\
49024.535 & $1$ & $216$ & $0$ & $195$ & Flash \\
49025.551 & $0$ & $931$ & $-0$ & $005$ & Flash \\
49026.535 & $1$ & $080$ & $0$ & $237$ & Flash \\
49027.543 & $1$ & $704$ & $0$ & $334$ & Flash \\
49028.543 & $1$ & $838$ & $0$ & $347$ & Flash \\
49029.535 & $1$ & $325$ & $0$ & $324$ & Flash \\
49030.543 & $0$ & $447$ & $0$ & $065$ & Flash \\
49031.539 & $-0$ & $062$ & $-0$ & $013$ & Flash \\
49032.562 & $-0$ & $713$ & $-0$ & $007$ & Flash \\
49033.555 & $-1$ & $587$ & $-0$ & $080$ & Flash \\
49034.578 & $-0$ & $853$ & $-0$ & $021$ & Flash \\
49035.547 & $-0$ & $591$ & $0$ & $185$ & Flash \\
49036.543 & $0$ & $119$ & $0$ & $068$ & Flash \\
49037.535 & $0$ & $395$ & $0$ & $152$ & Flash \\
49038.609 & $1$ & $029$ & $0$ & $295$ & Flash \\
49039.539 & $0$ & $918$ & $0$ & $181$ & Flash \\
49040.523 & $0$ & $932$ & $0$ & $411$ & Flash \\
49041.527 & $1$ & $068$ & $0$ & $080$ & Flash \\
49042.527 & $1$ & $476$ & $0$ & $410$ & Flash \\
49043.535 & $1$ & $903$ & $0$ & $278$ & Flash \\
49044.535 & $1$ & $838$ & $0$ & $321$ & Flash \\
49045.527 & $0$ & $906$ & $0$ & $375$ & Flash \\
49046.523 & $-0$ & $133$ & $0$ & $119$ & Flash \\
49047.527 & $-0$ & $710$ & $-0$ & $050$ & Flash \\
49048.523 & $-0$ & $807$ & $-0$ & $155$ & Flash \\
49049.520 & $-1$ & $262$ & $-0$ & $028$ & Flash \\
49050.523 & $-0$ & $731$ & $0$ & $198$ & Flash \\
49051.547 & $-0$ & $159$ & $0$ & $130$ & Flash \\
49052.547 & $0$ & $128$ & $0$ & $124$ & Flash \\
49053.523 & $0$ & $543$ & $0$ & $320$ & Flash \\
49054.512 & $0$ & $881$ & $0$ & $271$ & Flash \\
49055.516 & $1$ & $184$ & $0$ & $223$ & Flash \\
49056.512 & $1$ & $164$ & $0$ & $360$ & Flash \\
49057.520 & $1$ & $189$ & $0$ & $232$ & Flash \\
49058.512 & $1$ & $804$ & $0$ & $073$ & Flash \\
49059.512 & $1$ & $744$ & $0$ & $437$ & Flash \\
49060.516 & $1$ & $201$ & $0$ & $360$ & Flash \\
49061.512 & $0$ & $330$ & $0$ & $069$ & Flash \\
49062.512 & $-0$ & $417$ & $-0$ & $018$ & Flash \\
49063.516 & $-0$ & $959$ & $-0$ & $159$ & Flash \\
49065.512 & $-1$ & $023$ & $0$ & $069$ & Flash \\
49066.578 & $-0$ & $623$ & $0$ & $060$ & Flash \\
49067.512 & $0$ & $153$ & $0$ & $243$ & Flash \\
49068.512 & $0$ & $561$ & $0$ & $115$ & Flash \\
49069.516 & $0$ & $808$ & $0$ & $204$ & Flash \\
49070.516 & $1$ & $151$ & $0$ & $173$ & Flash \\
49071.516 & $1$ & $002$ & $0$ & $255$ & Flash \\
49072.516 & $1$ & $033$ & $0$ & $245$ & Flash \\
49073.527 & $1$ & $489$ & $0$ & $415$ & Flash \\
49074.520 & $1$ & $874$ & $0$ & $523$ & Flash \\
49075.520 & $1$ & $670$ & $0$ & $401$ & Flash \\
49076.516 & $0$ & $798$ & $0$ & $214$ & Flash \\
49078.520 & $-0$ & $707$ & $-0$ & $138$ & Flash \\
49079.516 & $-0$ & $971$ & $-0$ & $036$ & Flash \\
49080.512 & $-0$ & $912$ & $0$ & $014$ & Flash \\
49081.520 & $-0$ & $603$ & $0$ & $081$ & Flash \\
49082.516 & $-0$ & $067$ & $0$ & $071$ & Flash \\
49083.508 & $0$ & $429$ & $0$ & $246$ & Flash \\
49084.504 & $0$ & $825$ & $0$ & $023$ & Flash \\
49085.508 & $1$ & $124$ & $0$ & $353$ & Flash \\
49086.504 & $0$ & $960$ & $0$ & $034$ & Flash \\
49087.500 & $1$ & $122$ & $0$ & $273$ & Flash \\
49088.508 & $1$ & $287$ & $0$ & $391$ & Flash \\
49089.520 & $1$ & $648$ & $0$ & $365$ & Flash \\
49090.504 & $1$ & $803$ & $0$ & $331$ & Flash \\
49095.504 & $-1$ & $018$ & $-0$ & $115$ & Flash \\
49098.539 & $0$ & $115$ & $0$ & $124$ & Flash \\
49100.504 & $0$ & $947$ & $0$ & $232$ & Flash \\
49101.504 & $1$ & $112$ & $0$ & $299$ & Flash \\
49102.496 & $1$ & $024$ & $0$ & $386$ & Flash \\
49103.496 & $1$ & $242$ & $0$ & $341$ & Flash \\
49104.508 & $1$ & $612$ & $0$ & $477$ & Flash \\
49105.504 & $$ & $$ & $0$ & $472$ & Flash \\
49107.496 & $0$ & $485$ & $0$ & $056$ & Flash \\
49108.492 & $0$ & $284$ & $0$ & $156$ & Flash \\
49109.504 & $-1$ & $079$ & $-0$ & $215$ & Flash \\
49112.508 & $-0$ & $737$ & $0$ & $197$ & Flash \\
49387.544 & $-0$ & $912$ & $-0$ & $093$ & Flash \\
49392.566 & $0$ & $592$ & $0$ & $136$ & Flash \\
49395.539 & $1$ & $000$ & $0$ & $215$ & Flash \\
49401.526 & $-0$ & $235$ & $-0$ & $009$ & Flash \\
49406.519 & $-0$ & $063$ & $0$ & $055$ & Flash \\
49413.517 & $1$ & $850$ & $0$ & $385$ & Flash \\
49422.549 & $0$ & $007$ & $0$ & $097$ & Flash \\
49429.517 & $1$ & $437$ & $0$ & $386$ & Flash \\
49433.520 & $-0$ & $766$ & $-0$ & $120$ & Flash \\
49759.566 & $-0$ & $620$ & $0$ & $012$ & Heros \\
49771.591 & $0$ & $055$ & $-0$ & $034$ & Heros \\
49776.570 & $0$ & $161$ & $0$ & $037$ & Heros \\
49777.574 & $0$ & $499$ & $0$ & $123$ & Heros \\
49778.519 & $0$ & $831$ & $0$ & $132$ & Heros \\
49779.518 & $1$ & $146$ & $0$ & $146$ & Heros \\
49780.518 & $1$ & $301$ & $0$ & $223$ & Heros \\
49781.517 & $1$ & $300$ & $0$ & $160$ & Heros \\
49782.520 & $1$ & $511$ & $0$ & $243$ & Heros \\
49783.517 & $1$ & $945$ & $0$ & $270$ & Heros \\
49784.517 & $1$ & $851$ & $0$ & $305$ & Heros \\
49785.517 & $1$ & $475$ & $0$ & $272$ & Heros \\
49786.520 & $0$ & $472$ & $0$ & $021$ & Heros \\
49787.543 & $-0$ & $473$ & $-0$ & $109$ & Heros \\
49788.522 & $-0$ & $665$ & $-0$ & $210$ & Heros \\
49789.518 & $-0$ & $940$ & $-0$ & $035$ & Heros \\
49790.505 & $-0$ & $848$ & $-0$ & $156$ & Heros \\
49791.545 & $-0$ & $074$ & $0$ & $025$ & Heros \\
49792.518 & $0$ & $433$ & $0$ & $056$ & Heros \\
49793.510 & $0$ & $687$ & $0$ & $090$ & Heros \\
49794.506 & $1$ & $034$ & $0$ & $099$ & Heros \\
49795.509 & $1$ & $302$ & $-0$ & $020$ & Heros \\
49796.502 & $1$ & $266$ & $0$ & $125$ & Heros \\
49797.505 & $1$ & $439$ & $0$ & $205$ & Heros \\
49798.498 & $1$ & $767$ & $0$ & $244$ & Heros \\
50030.595 & $1$ & $927$ & $0$ & $417$ & Elodie \\
50716.650 & $-0$ & $568$ & $0$ & $157$ & Heros \\
50717.617 & $0$ & $047$ & $0$ & $154$ & Heros \\
50718.589 & $0$ & $388$ & $0$ & $192$ & Heros \\
50719.612 & $0$ & $760$ & $0$ & $256$ & Heros \\
50726.583 & $1$ & $150$ & $0$ & $298$ & Heros \\
50727.597 & $0$ & $158$ & $$ & $$ & Heros \\
51097.892 & $-0$ & $034$ & $0$ & $051$ & Feros \\
51134.880 & $0$ & $300$ & $0$ & $131$ & Feros \\
51145.866 & $-0$ & $823$ & $-0$ & $077$ & Feros \\
51148.801 & $-0$ & $425$ & $0$ & $103$ & Feros \\
51151.776 & $0$ & $891$ & $0$ & $228$ & Feros \\
51172.579 & $1$ & $843$ & $0$ & $374$ & Feros \\
51177.579 & $-0$ & $963$ & $-0$ & $018$ & Feros \\
51179.593 & $-0$ & $568$ & $0$ & $099$ & Feros \\
51180.513 & $0$ & $358$ & $0$ & $145$ & Feros \\
51191.700 & $-0$ & $777$ & $-0$ & $053$ & Feros \\
51192.551 & $-0$ & $956$ & $-0$ & $049$ & Feros \\
51373.931 & $1$ & $343$ & $0$ & $401$ & Feros \\
51380.936 & $-0$ & $011$ & $0$ & $150$ & Feros \\
51381.937 & $0$ & $459$ & $0$ & $188$ & Feros \\
51382.939 & $0$ & $920$ & $0$ & $184$ & Feros \\
51383.932 & $1$ & $147$ & $0$ & $249$ & Feros \\
51384.928 & $1$ & $094$ & $0$ & $236$ & Feros \\
51385.935 & $1$ & $083$ & $0$ & $191$ & Feros \\
51386.946 & $1$ & $406$ & $0$ & $318$ & Feros \\
51389.933 & $0$ & $892$ & $0$ & $295$ & Feros \\
51390.954 & $-0$ & $227$ & $0$ & $050$ & Feros \\
51391.947 & $-0$ & $934$ & $-0$ & $083$ & Feros \\
51392.900 & $-1$ & $120$ & $-0$ & $073$ & Feros \\
51393.950 & $-1$ & $265$ & $-0$ & $007$ & Feros \\
51394.935 & $-1$ & $140$ & $0$ & $023$ & Feros \\
53329.563 & $0$ & $979$ & $0$ & $225$ & Elodie \\
51578.422 & $-1$ & $242$ & $-0$ & $103$ & Musicos \\
51578.454 & $-1$ & $407$ & $-0$ & $135$ & Musicos \\
51579.393 & $-1$ & $289$ & $-0$ & $061$ & Musicos \\
51579.424 & $-1$ & $222$ & $-0$ & $056$ & Musicos \\
51579.454 & $-1$ & $095$ & $-0$ & $041$ & Musicos \\
51587.374 & $1$ & $493$ & $0$ & $205$ & Musicos \\
51590.382 & $1$ & $280$ & $0$ & $243$ & Musicos \\
51590.413 & $1$ & $103$ & $0$ & $249$ & Musicos \\
51596.391 & $-0$ & $074$ & $0$ & $177$ & Musicos \\
51596.421 & $-0$ & $086$ & $0$ & $123$ & Musicos \\
51596.452 & $0$ & $056$ & $0$ & $136$ & Musicos \\
51597.351 & $0$ & $332$ & $0$ & $161$ & Musicos \\
51597.382 & $0$ & $405$ & $0$ & $193$ & Musicos \\
51597.413 & $0$ & $412$ & $0$ & $190$ & Musicos \\
51597.444 & $0$ & $490$ & $0$ & $168$ & Musicos \\
51599.353 & $1$ & $196$ & $0$ & $219$ & Musicos \\
51599.384 & $1$ & $230$ & $0$ & $216$ & Musicos \\
51599.416 & $1$ & $331$ & $0$ & $214$ & Musicos \\
51600.346 & $1$ & $483$ & $0$ & $247$ & Musicos \\
51600.378 & $1$ & $474$ & $0$ & $223$ & Musicos \\
51600.411 & $1$ & $386$ & $0$ & $236$ & Musicos \\
51601.361 & $1$ & $313$ & $0$ & $220$ & Musicos \\
51601.392 & $1$ & $379$ & $0$ & $209$ & Musicos \\
51601.423 & $1$ & $337$ & $0$ & $219$ & Musicos \\
51602.349 & $1$ & $534$ & $0$ & $275$ & Musicos \\
51602.388 & $1$ & $551$ & $0$ & $266$ & Musicos \\
51602.419 & $1$ & $498$ & $0$ & $247$ & Musicos \\
51603.377 & $2$ & $000$ & $0$ & $316$ & Musicos \\
51603.408 & $1$ & $964$ & $0$ & $322$ & Musicos \\
51603.439 & $2$ & $072$ & $0$ & $283$ & Musicos \\
51606.353 & $0$ & $803$ & $0$ & $258$ & Musicos \\
51606.384 & $0$ & $797$ & $0$ & $211$ & Musicos \\
51606.416 & $0$ & $637$ & $0$ & $224$ & Musicos \\
51608.370 & $-0$ & $521$ & $0$ & $057$ & Musicos \\
51608.400 & $-0$ & $542$ & $0$ & $033$ & Musicos \\
51608.431 & $-0$ & $583$ & $0$ & $012$ & Musicos \\
51609.360 & $-0$ & $734$ & $0$ & $121$ & Musicos \\
51609.391 & $-0$ & $705$ & $0$ & $130$ & Musicos \\
51609.422 & $-0$ & $779$ & $0$ & $198$ & Musicos \\
53740.534 & $-0$ & $271$ & $0$ & $157$ & Feros \\
53744.891 & $1$ & $087$ & $0$ & $215$ & Espadons \\
54169.854 & $-0$ & $886$ & $0$ & $004$ & Espadons \\
54456.748 & $1$ & $798$ & $0$ & $389$ & Espadons \\
54457.748 & $1$ & $952$ & $0$ & $454$ & Espadons \\
54505.248 & $1$ & $370$ & $$ & $$ & Lhires III,1 \\
54505.375 & $1$ & $110$ &  $$ & $$ &Lhires III,3  \\
54507.262 & $-0$ & $462$ &  $$ & $$ & Lhires III,1 \\
54508.264 & $-0$ & $765$ &  $$ & $$ & Lhires III,1 \\
54509.248 & $-1$ & $02$ &  $$ & $$ & Lhires III,1  \\
54510.259 & $-1$ & $327$ &  $$ & $$ & Lhires III,1  \\
54512.244 & $-0$ & $273$ &  $$ & $$ & Lhires III,1  \\
54513.28 & $0$ & $706$ &  $$ & $$ & Lhires III,1  \\
54514.245 & $0$ & $792$ &  $$ & $$ & Lhires III,1  \\
54515.253 & $0$ & $907$ &  $$ & $$ & Lhires III,1  \\
54516.257 & $0$ & $836$ &  $$ & $$ & Lhires III,1  \\
54521.248 & $0$ & $614$ &  $$ & $$ & Lhires III,1  \\
54527.294 & $0$ & $039$ &  $$ & $$ & Lhires III,2  \\
54529.272 & $0$ & $952$ &  $$ & $$ & Lhires III,2  \\
54531.252 & $1$ & $426$ &  $$ & $$ & Lhires III,2  \\
54544.297 & $0$ & $802$ &  $$ & $$ & Lhires III.2  \\
\hline
\hline
\label{tab:eqw2}
\end{longtable}

\end{document}